\documentclass[prx,twocolumn,preprintnumbers,amsmath,amssymb,showpacs]{revtex4}
\usepackage{graphicx, bm}
\usepackage{subfigure,amssymb}
\usepackage{bbold}
\usepackage{color}

\begin{document}
\title{\bf Disorder driven itinerant quantum criticality of three dimensional massless Dirac fermions}
\author{J. H. Pixley}
\affiliation{Condensed Matter Theory Center and Joint Quantum Institute, Department of Physics, University of Maryland, College Park, Maryland 20742- 4111 USA}
\author{Pallab Goswami}
\affiliation{Condensed Matter Theory Center and Joint Quantum Institute, Department of Physics, University of Maryland, College Park, Maryland 20742- 4111 USA}
\author{S. Das Sarma}
\affiliation{Condensed Matter Theory Center and Joint Quantum Institute, Department of Physics, University of Maryland, College Park, Maryland 20742- 4111 USA}

\date{\today}
\begin{abstract}
Progress in the understanding of quantum critical properties of itinerant electrons has been hindered by the lack of effective models which are amenable to controlled analytical and numerically exact calculations. Here we establish that the disorder driven semimetal to metal quantum phase transition of three dimensional massless Dirac fermions could serve as a paradigmatic toy model for studying itinerant quantum criticality, which is solved in this work by exact numerical and approximate field theoretic calculations. As a result, we establish the robust existence of a non-Gaussian universality class, and also construct the relevant low energy effective field theory that could guide the understanding of quantum critical scaling for many strange metals. Using the kernel polynomial method (KPM), we provide numerical results for the calculated dynamical exponent ($z$) and correlation length exponent ($\nu$) for the disorder-driven semimetal (SM) to diffusive metal (DM) quantum phase transition at the Dirac point for several types of disorder, establishing its universal nature and obtaining the numerical scaling functions in agreement with our field theoretical analysis.

\end{abstract}

\pacs{71.10.Hf,72.80.Ey,73.43.Nq,72.15.Rn}

\maketitle

\section{Introduction} Phase transitions are ubiquitous in the natural world ranging from the solidification of water to the thermal suppression of magnetism in iron or superconductivity in metals.  Through the construction of simplified effective ``toy'' models, such as the Ising model to describe classical magnets~\cite{Goldenfeld-book}, generic theories can be developed to gain invaluable physical insights
and
explain the universal features of experimental data across vastly different systems. Due to the quantum mechanical zero-point motion, phase transitions can also occur at absolute zero temperature~\cite{Sachdev-book} driven entirely by quantum (rather than thermal) fluctuations, which can be accessed by tuning a non-thermal control parameter $g$, such as disorder, doping, magnetic field, or pressure. When such a quantum phase transition (QPT) is continuous, the quantum critical point (QCP) (located at a critical coupling $g_c$ akin to the critical temperature for thermal phase transitions) exhibits scale invariance, and the universal scaling properties of physical quantities are governed by the divergent length ($\xi_l$)  and time ($\xi_t$)  scales
$$\xi_l  \sim |g-g_c|^{-\nu}, \: \: \xi_t \sim \xi_l^z \sim |g-g_c|^{-\nu z},$$
where the critical exponents $\nu$ and $z$
characterize the correlations in space and time respectively. These notions have been put on a solid theoretical foundation, e.g. by adding non-trivial quantum dynamics into the Ising model~\cite{Young-1975}, which has allowed for a simplified setting to understand magnetic quantum phase transitions in general.  The critical exponents $\nu$ and $z$ define the universality class of the quantum phase transition, and critical scaling functions connecting various physical parameters define the nature of the QCP.

\begin{figure}[htbp]
  \centering
  \includegraphics[width=0.9\linewidth]{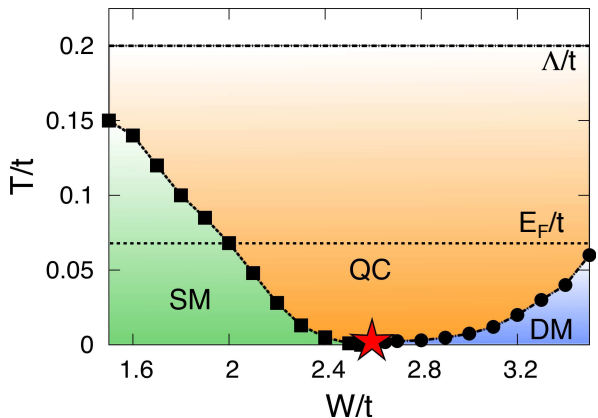}
\caption{(Color Online) Numerically calculated finite temperature $(T)$ phase diagram as a function of disorder $(W)$ computed for a linear system size $L=110$ (with $t$, the band hopping energy, setting the energy scale). The Dirac semimetal (SM), the diffusive metal (DM) and the QCP separating them only exist at zero temperature and there is no finite temperature phase transition. The squares mark the cross over out of the SM phase (green) into the quantum critical (QC) region (orange), which is anchored by the QCP at the critical disorder value $W_c/t=2.55\pm0.05$ (star).  This cross over is determined by where the specific heat fails to be described by $C_V\sim T^3$.  The circles mark the cross over out of the DM phase (blue) into the QC region, which is marked by where the specific heat fails to be described by $C_V \sim T$. The dashed lines connecting the squares/circles are a guide to the eye.  The dashed line at $\Lambda/t$ marks the high energy cut off above which the low energy continuum description in terms of massless Dirac fermions is inapplicable. In a narrow region inside the quantum critical regime the specific heat manifests non-Fermi-liquid behavior $C_V \sim T^2$ which gradually crosses over to the power law behaviors of SM and DM. The dashed horizontal line $E_F/t$ is a schematic for the (generally unknown) Fermi energy associated with any residual doping that drives the system slightly away from the precise Dirac point, indicating the low-energy cut off above which the zero-doping Dirac point theory is applicable.  Note that as long as $E_F$ is not too large, quantum criticality manifests itself in the orange region of the phase diagram, even though the QCP itself (the star at $W_c$ and $T=0$) is `hidden' by the residual doping. At temperature and energy scales smaller than $E_F$, the system behaves as a conventional metal or diffusive Fermi liquid. In a nominally undoped system, the inequality $\Lambda \gg E_F$ applies, and quantum criticality is observable for $k_BT>E_F$.
}
 \label{fig:pd}

\end{figure}

Precisely at the QCP ($g=g_c$) and at $T=0$, one has true scale invariance as the characteristic correlation lengths ($\xi_l$ and $\xi_t$) are infinite. Even though the QCP occurs strictly at zero temperature, its effects on physical properties manifest over a broad range of temperatures ($T$) and energies ($E$), such that
$$E, \, k_BT \gg \xi^{-1}_{t}, $$
where $k_B$ is the Boltzmann constant.  
This critical region, known as the quantum critical fan (see Fig. 1), is naturally accessible in the laboratory and is widely studied~\cite{Sondi-1997,Giamarchi-2008,Si-2010}. The generic magnetic QCPs for most insulating systems can be simply understood in terms of the Landau-Ginzburg-Wilson order parameter theory in $(d+z)$ dimensions, since there is only one gapless bosonic degree of freedom, namely the order parameter. By contrast, at the the metallic or itinerant QCPs there are at least two sets of gapless degree of freedom, (i) the itinerant fermions and (ii) the bosonic order parameter, whose mass has been tuned to zero (at $g=g_c$). Thus, to construct a successful description of a metallic or itinerant QCP, it is necessary to address both types of gapless excitations on an equal footing, which is generally a challenging task.

In this work, we show that the disorder driven semimetal to diffusive metal (SM-DM) QCP of three-dimensional Dirac fermions is a quintessential toy model for gaining qualitatively new insights into the general nature of itinerant quantum criticality.
For a random scalar potential, this problem has been studied by mean-field~\cite{Fradkin-1986} and perturbative field theoretical calculations~\cite{Goswami-2011, Bitan-2014, Leo-2015,Sergey-2015, Sergey2-2015}. These analytical calculations show that above the lower critical dimensionality $d=2$, the average density of states (DOS) at zero energy vanishes inside the SM for sufficiently weak disorder, and it becomes finite beyond a critical disorder strength leading to a diffusive metal (DM). 
There is no such finite disorder phase transition in two dimensions (e.g. graphene), where the system becomes a diffusive metal already for infinitesimal disorder.
Through a $d=2+\epsilon $ expansion of the critical properties, it is found that the leading order results for the critical exponents in $d=3$ are $z=3/2$ and $\nu=1$, for both potential and axial disorder, while there should be no such transition for mass disorder~\cite{Goswami-2011}. Recently, numerical calculations have also come to bear on the problem through the study of a three dimensional disordered topological insulator~\cite{Kobayashi-2013,Kobayashi-2014}, a three-dimensional layered Chern insulator~\cite{Liu-2015}, a Weyl semimetal~\cite{Brouwer-2014,Sbierski-2015,Bera-2015}, and the phase diagram of Dirac~\cite{Pixley-2015} and Weyl~\cite{Shapourian-2015} semimetals. For the case of a single Weyl cone which can be realized on the surface of a four dimensional topological insulator, the critical exponents have been numerically obtained to high accuracy~\cite{Sbierski-2015}. 
For avoiding any misunderstanding, we mention that there is a second quantum phase transition~\cite{Pixley-2015} for stronger disorder ($W_l \gg W_c$ than considered in Fig.~\ref{fig:pd}), where the diffusive metal undergoes an Anderson localization. In this work we do not address the Anderson localization transition at all and exclusively focus on the SM-DM QPT restricting ourselves to disorder strengths below the threshold for the Anderson transition~\cite{Pixley-2015}.

This itinerant QPT is addressed in our work through numerically exact calculations on sufficiently large system sizes ($\sim10^6$ lattice points) and approximate field theoretic analysis, which we use to establish:
\begin{enumerate}

\item The numerical value 
of $z$ is universal, being independent of the disorder type and the details of their probability distribution (box versus a Gaussian distribution) and we obtain numerically $z=1.46 \pm 0.05$ (in good agreement with other numerical studies on different models~\cite{Kobayashi-2014,Sbierski-2015,Liu-2015}). For  the correlation length exponent we cannot pinpoint a universal value and we find a value of $\nu$ that lies in the range $\sim 0.9-1.5$. (Our numerical technique, while providing a fairly precise value of $z$, is relatively imprecise in obtaining $\nu$ because of the uncertainties in the precise determination of the critical disorder strength $W_c$ which turns out to be crucial in calculating $\nu$, but not $z$.)

\item The existence of a broad quantum critical fan at finite temperatures (see Fig. 1), where the fan occupies a large parameter space.

\item We find considerable agreement between the numerically exact crossover scaling functions and their approximate analytical forms (Appendix A) for the average density of states and the specific heat. The approximate analytical calculations are based on one loop renormalization group analysis which predicts $z=3/2$ and $\nu=1$.

\item We also construct an order parameter field theory coupled to itinerant fermions, which illustrates how the notion of well defined quasiparticle excitations can break down in the quantum critical regime. The microscopic model we solve is noninteracting (with disorder as the tuning parameter for the QCP), and the sample to sample fluctuations of disorder give rise to effective electronic interactions of a statistical nature. Even though this does not pertain to correlated electronic systems \emph{per se}, the theoretical concepts, regarding the itinerant quantum critical scaling properties that we are able to establish, should be generally applicable to metallic quantum critical points. In particular, our theoretical finding and the numerical verification of the non-Fermi-liquid behavior and crossover scaling  functions in the quantum critical regime should have generic qualitative applicability to itinerant quantum critical systems. Therefore, this model may effectively serve as an ``Ising model'' for itinerant quantum criticality for future theoretical work.

\item Our theoretical results are directly pertinent for describing the recently discovered gapless Dirac semiconductors or semimetals~\cite{Neupane-2014,Borisenko-2014,Liu-2014,Liu2-2014,Xu-2015}, where the valence and conduction bands touch linearly at isolated points in the Brillioun zone. In particular, we expect that the quantum critical fan established in this work will be experimentally observable in Dirac semimetals with a very low carrier concentration. We therefore propose Na$_3$Bi, which has a small Fermi energy~\cite{Liu2-2014}, as one of the most promising candidate materials for realizing the predicted critical properties. Our results can also be relevant for the three dimensional topological quantum phase transitions between a topological and band insulator in the presence of disorder~\cite{Teo-2008,Xu-2011,Sato-2011,Brahlek-2012,Wu-2013}. Although our quantum critical results strictly apply only for the intrinsic (i.e. undoped) Dirac semimetal, the results should remain valid even for finite doping as long as the doping density is low enough so that the associated chemical potential is smaller than the temperature~\cite{DasSarma-2015}.
    
In our numerically calculated phase diagram (Fig.~\ref{fig:pd}) we depict the SM, DM, and the critical fan regions with green, blue, and orange colors respectively whereas the (bottom) dashed horizontal line schematically indicates the Fermi level (or chemical potential) associated with any residual (unintentional and hence unknown) doping in the system providing the low-energy cut off for the quantum critical theory.  The high-energy cut off is indicated by the top horizontal line which is associated with the usual ultraviolet cut off for studying Dirac physics on a lattice.  We note that the invariable presence of some residual doping in the system producing  the low-energy cut-off $E_F$ in Fig.~\ref{fig:pd} acts to `hide' the QCP by introducing an incipient (unintentional) metallic phase.  Although the QCP itself is inaccessible and hidden, its effect persists in the quantum critical regime (the orange region above the $E_F$ line in Fig.~\ref{fig:pd}) provided that the doping is not too high.  This situation is not dissimilar to many examples of itinerant quantum criticality in correlated metallic systems. 

\end{enumerate}

The remainder of the paper is organized as follows: In section~\ref{sec:model} we introduce the lattice model and its continuum limit. 
 In section~\ref{sec:numerics} we present detailed numerical results, which is followed by the construction of the order parameter field theory in section~\ref{sec:orderparameter}, and we conclude in section~\ref{sec:conclusions}. In Appendix~\ref{sec:appendix-A}, we theoretically determine the general scaling formulas used to analyze the numerical data.  Finally additional numerical details for our calculations are provided in Appendix~\ref{sec:appendix-C}.

\section{Lattice Model and continuum limit}
\label{sec:model}
We begin by introducing the lattice model and its continuum limit that will be the focus of this work. We consider non-interacting electrons hopping on a simple cubic lattice with periodic boundary conditions in the presence of a random on site potential. The Hamiltonian is
\begin{eqnarray}
H=\frac{1}{2}\sum_{{\bf r},j}\left(i t \psi_{{\bf r}}^{\dag}\,\alpha_{j}\psi_{{\bf r}+{\bf \hat{e}}_{j}}  + \mathrm{H.c}\right)+\sum_{{\bf r}} V({\bf r})\psi_{{\bf r}}^{\dag}A_W\psi_{{\bf r}} \,\,\,\,\,\,
\label{eqn:ham}
\end{eqnarray}
where $\psi_{{\bf r}}$ is a four component spinor. For non-superconducting systems with conserved electric charge we can choose $\psi_{{\bf r}}^T=(c_{{\bf r},+,\uparrow},c_{{\bf r},-,\uparrow},c_{{\bf r},+,\downarrow},c_{{\bf r},-,\downarrow})$ which is composed of an electron annihilation operator $c_{{\bf r},\tau,s}$ at site ${\bf r}$, with parity  $\tau=\pm$, and spin-projections $s=\uparrow/\downarrow$. We work in the Dirac representation and therefore the matrices are
\begin{equation}
\alpha_j = \left( \begin{array}{cc}
0 & \sigma_j  \\
\sigma_j & 0 \end{array} \right),
\,
\gamma_5 =  \left( \begin{array}{cc}
0 & \bm{1}  \\
\bm{1} & 0 \end{array} \right),
\,
\beta =  \left( \begin{array}{cc}
\bm{1} & 0  \\
0 & -\bm{1} \end{array} \right),
\end{equation}
where $\sigma_j$ denotes the Pauli matrices and $\bm{1}$ denotes the $2\times2$ identity matrix.
Each site is labeled by the vector ${\bf r}$ and the nearest neighbors are connected by the unit vectors ${\bf \hat{e}}_{j}$, with $j=1,2,3$. The strength of hopping between neighboring sites is determined by $t$, and the random potential is $V({\bf r})$. (We often use $t=1$ in the rest of this paper using the hopping as the unit of energy in the problem.) To study the universality of the transition we consider two types of random distributions for $V({\bf r})$: (i)  a box distribution (BD), i.e. a randomly distributed variable between $[-W/2,W/2]$, (ii) a Gaussian distribution (GD) with zero mean and variance $W^2$. The type of disorder is specified by the $A_W$ matrix, we consider three different types of disorder: (1) axial disorder $A_5 = \gamma_5$, (2) mass disorder $A_m = \beta$, and (3) potential disorder $A_p = I_{4\times4}$ (where $ I_{4\times4}$ denotes the four by four identity matrix). Physically, each type of disorder can naturally occur in experimental systems without any control over which one may appear or produce the dominant effect, and it is not unreasonable to assume that the QPT tuned by each type of disorder belongs to a different universality class (which turns out not to be the case here as we will see).

\begin{figure*}[htbp]
\centering
\begin{minipage}{.5\textwidth}
  \centering
  \includegraphics[width=0.65\linewidth,angle=-90]{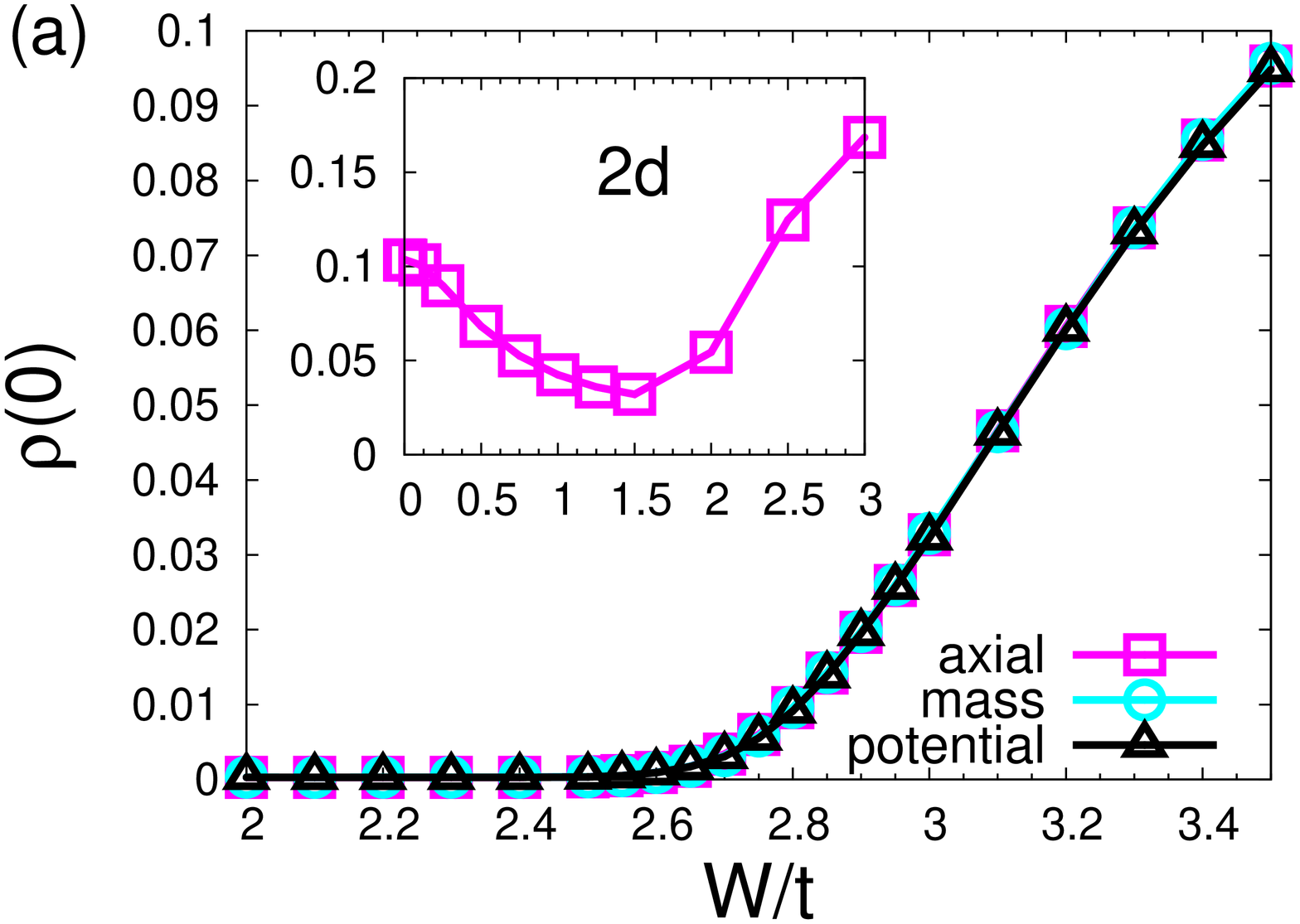}
\end{minipage}%
\begin{minipage}{.5\textwidth}
  \centering
  \includegraphics[width=0.9\linewidth,angle=0]{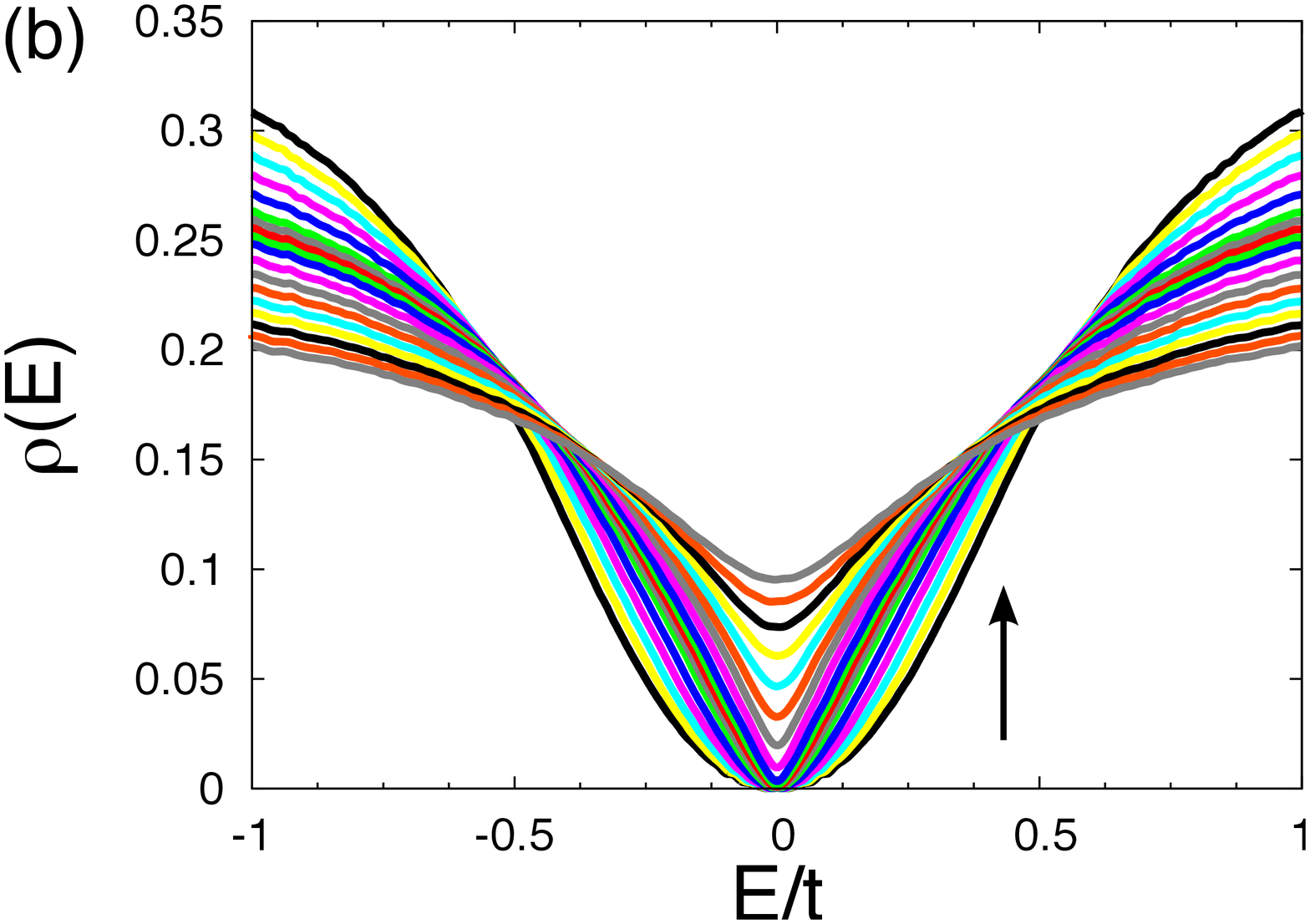}
 \end{minipage}
  \newline
 \centering
\begin{minipage}{.5\textwidth}
  \centering
  \includegraphics[width=0.7\linewidth,angle=-90]{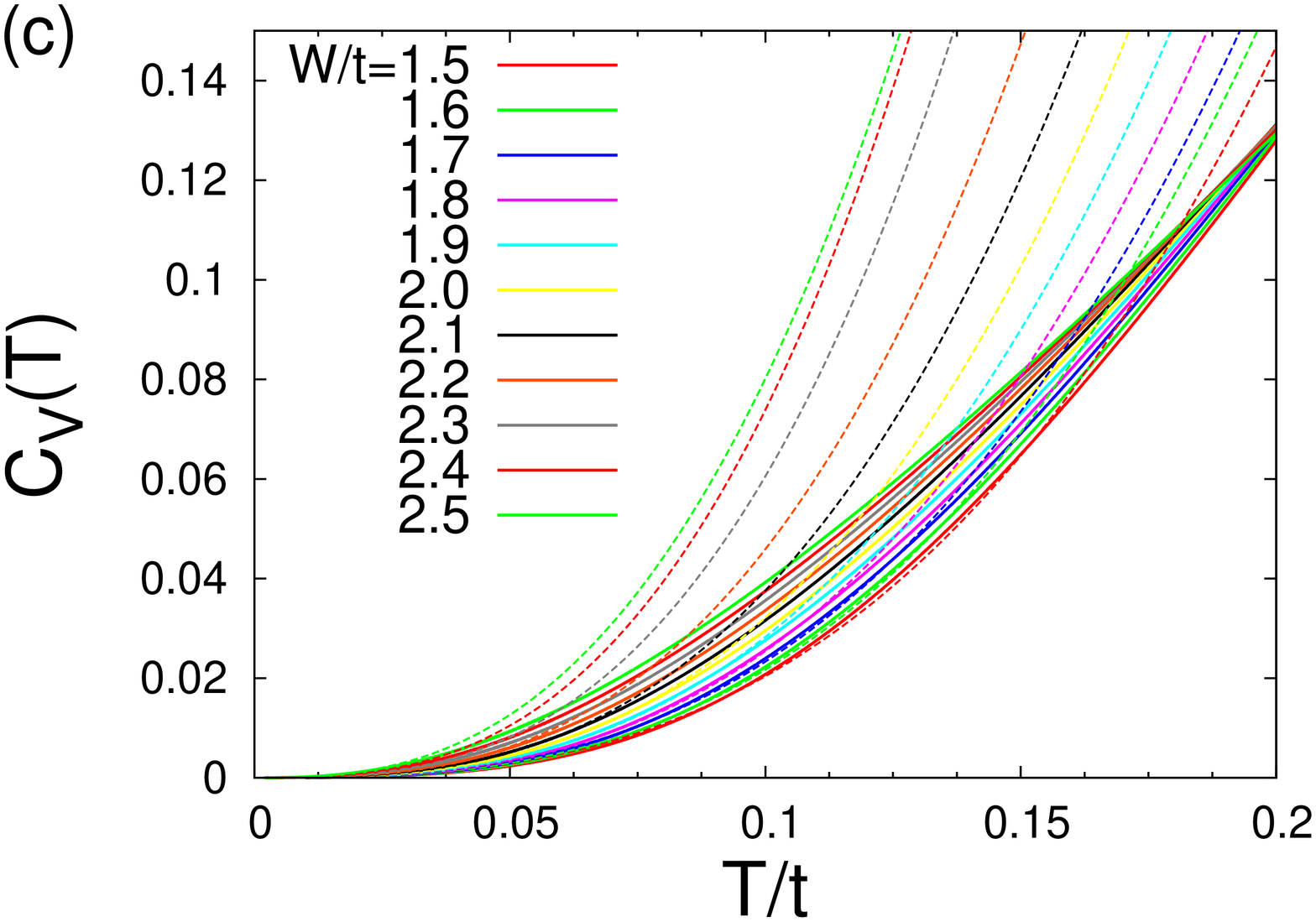}
\end{minipage}%
\begin{minipage}{.5\textwidth}
  \centering
  \includegraphics[width=0.7\linewidth,angle=-90]{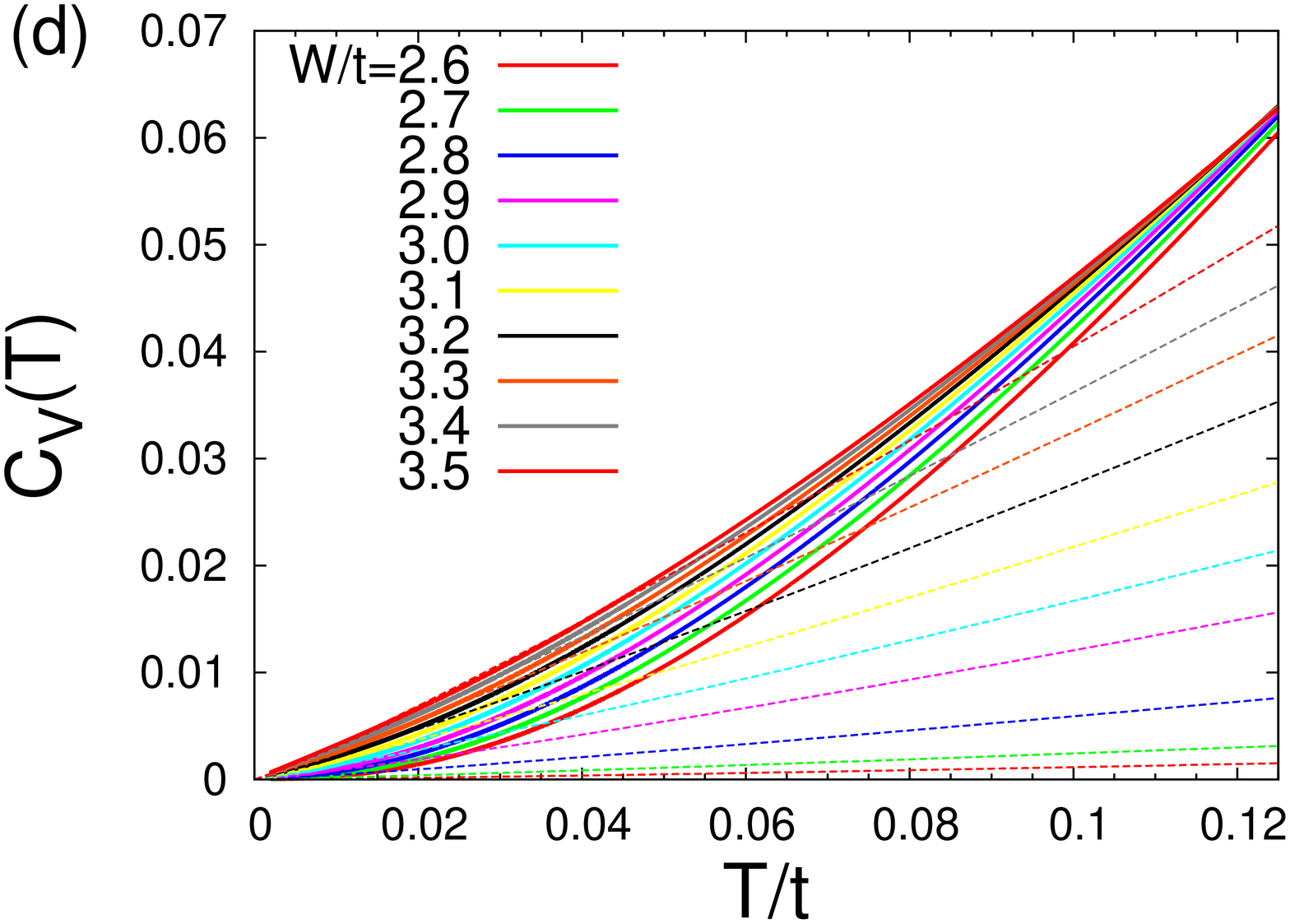}
 \end{minipage}
\caption{(Color Online) Numerically (KPM)-calculated DOS as a function of disorder strength for a linear system size $L=110$. (a) DOS at zero energy $\rho(0)$ as a function of disorder strength for each type of disorder matrix $A_W$, which shows that axial, mass, and potential disorder yield identical results. (Inset) Corresponding calculation for axial disorder in two dimensions [i.e. setting the hopping in the $z$-direction to zero in Eq.~(\ref{eqn:ham})] clearly establishing that there is no stable SM phase and $\rho(0)$ is finite even for an infinitesimal amount of disorder. (b) DOS for axial disorder as a function of energy passing through the QCP with the arrow indicating increasing disorder strength. Specific heat (solid lines) with axial disorder for $W< W_c$ (c) and $W>W_c$ (d), the fits (dashed lines) are to the leading $C_V \sim T^3$ in the SM and $C_V \sim T$ in the DM. The temperatures at which the fits no longer apply determine the cross over boundary shown in Fig.~\ref{fig:pd}.
}
\label{fig:2}
\end{figure*}

In addition to the global U(1) symmetry for total number conservation, the tight binding model also has a continuous U(1) chiral symmetry, as the Hamiltonian commutes with $ \sum_{{\bf r}} \psi_{{\bf r}}^{\dag} \gamma_5 \psi_{{\bf r}}$, where $\gamma_5=i\alpha_1\alpha_2\alpha_3$. This Hamiltonian serves as a model for a topological Dirac semimetal with eight Dirac cones at the high symmetry points of the cubic Brillouin zone, and the excitations around different cones may be coupled by intervalley scattering due to disorder. As a result of the U(1) chiral symmetry we expect the axial and potential disorders to have
identical critical properties, manifesting universality. An important question is whether the universality of the SM-DM QCP remains unaffected when the $U(1)$ chiral symmetry is absent.
We address this question by studying the effects of mass disorder, which reduces the U(1) chiral symmetry down to a $Z_2$ chiral or particle-hole symmetry, described by the relation $\{H, \sum_{{\bf r}} \psi_{{\bf r}}^{\dag} i\beta\gamma_5 \psi_{{\bf r}} \}=0$. While potential disorder is only relevant for non-superconducting systems, axial and mass disorder can naturally appear for superconducting systems. Due to the discrete particle-hole symmetry, the mass disorder problem can appear for class AIII or DIII (depending on the representation of the spinor). By contrast, the axial disorder displays the discrete particle hole symmetry with respect to both $\beta$, $i\beta \gamma_5$ and any arbitrary linear combination of them. Thus, the axial disorder model is shared by both AIII and DIII Altland-Zirnbauer classes. One of our completely unanticipated interesting new findings is that the average density of states at zero energy ($\rho(0)$) across the SM-DM QCP is independent of the type of disorder (potential, mass, or axial) driving the transition, see Fig.~\ref{fig:2} (a). As $\rho(0)$ acts as the order parameter of the SM-DM transition, this implies same power law dependence of $\rho(0)$ for potential, axial, and mass disorder. Consequently, the SM-DM QPTs driven by potential, mass, and axial disorders belong to the same universality class. We establish this universality of the disorder-tuned SM-DM transitions through nonperturbative exact numerical calculations.

In the absence of disorder, the model reduces to the well known Dirac Hamiltonian on a lattice, with a dispersion
\begin{equation}E_0({\bf k})=\pm t\sqrt{\sum_{j}\sin(k_{j})^2} \; . \end{equation}
Expanding the dispersion in the low energy limit gives rise to eight Dirac cones with linearly dispersing excitations around eight high symmetry points of the cubic Brillouin zone. Considering only the long wavelength degrees of freedom (which are the most dominant at a second order phase transition) in the presence of disorder we arrive at the following continuum Hamiltonian as the effective theory
\begin{equation}
\mathcal{H}_D= \psi^{\dagger}_{\lambda}\left(-i v \alpha_{j} \partial_{j}\otimes I_{\lambda \sigma}+ V({\bf x})A_W\otimes B_{\lambda \sigma}  \right)\psi_{\sigma},
\label{eqn:ham2}
\end{equation}
where $I_{\lambda\sigma}$ is an eight by eight identity matrix in the valley space, $B_{\lambda \gamma}$ is the intervalley mixing matrix,
 and now the Dirac spinor carries a valley or flavor index $\lambda$. In the absence of intervalley scattering there is an $SU(8)$ flavor symmetry. The intervalley scattering due to disorder breaks this flavor symmetry.  We mention, however, that for long-ranged disorder, e.g., Coulomb impurities, such a disorder-induced valley symmetry breaking may be weak and thus operational only at very low temperatures, which appears to be the situation in (the two-dimensional Dirac material) graphene.  In such a situation, the independent Dirac cone approximation that ignores intervalley scattering effects is valid down to very low temperatures.

For simplicity of the analysis, the analytic calculations are performed with the independent Dirac cone approximation, while neglecting intervalley scattering (i.e. $B_{\lambda \sigma} \rightarrow I_{\lambda\sigma}$). We choose a Gaussian white noise disorder distribution with a variance $\tilde{\Delta}$, and define the dimensionless disorder strength $\Delta=\tilde{\Delta}\Lambda^{d-2} \Omega_d/[(2\pi)^2v^2]$ where $\Lambda$ is the large momentum cut off and $\Omega_d=2\pi^{d/2}/\Gamma(d/2)$ is the area of the unit sphere in $d$ dimensions, and $\Gamma(x)$ is the gamma function. 
 Note that for analytic calculations we refer to the reduced distance from the QCP as $\delta=|\Delta-\Delta_c|/\Delta_c$, while for the numerical work we denote the reduced distance as $\delta=|W-W_c|/W_c$. To avoid confusion we explicitly mention our definitions when we use $\delta$.
The one loop calculation (or to the lowest order in $\epsilon$) for potential or axial disorder yields the same beta function~\cite{Goswami-2011} 
and the SM-DM QPT is controlled by the QCP located at $\Delta_c=(d-2)/2=\epsilon/2$ with $z=1+\epsilon/2$ and $\nu=1/(d-2)$. For $d=3$, this leads to $\nu=1$, and $z=3/2$. The implications of one loop calculations for determining quantum critical scaling properties are discussed in detail in Appendix~\ref{sec:appendix-A}.
By contrast, the one-loop beta function shows intravalley mass disorder to be an irrelevant perturbation and does not predict any QPT. However, through our nonperturbative numerical calculations we show that the mass disorder with intervalley scattering actually drives a SM-DM QPT (see Fig.~\ref{fig:2} (a)), in contrast to the one loop perturbative result with only intravalley scattering.

The important question is whether the results derived at the one loop level for potential and axial disorders in the single-cone approximation remains valid in the presence of intervalley scattering, i.e.,  how robust the analytically obtained values of the critical exponents and scaling functions in the general situation are where the precise theoretical assumptions necessary for obtaining the analytical results may no longer apply.
To answer these questions non-perturbatively, we have performed detailed numerical calculations on the tight binding model in Eq.~(\ref{eqn:ham}) using large system sizes, which we present in the next section.

\begin{figure*}[htb]
\centering
\begin{minipage}{.5\textwidth}
  \centering
  \includegraphics[width=0.7\linewidth,angle=-90]{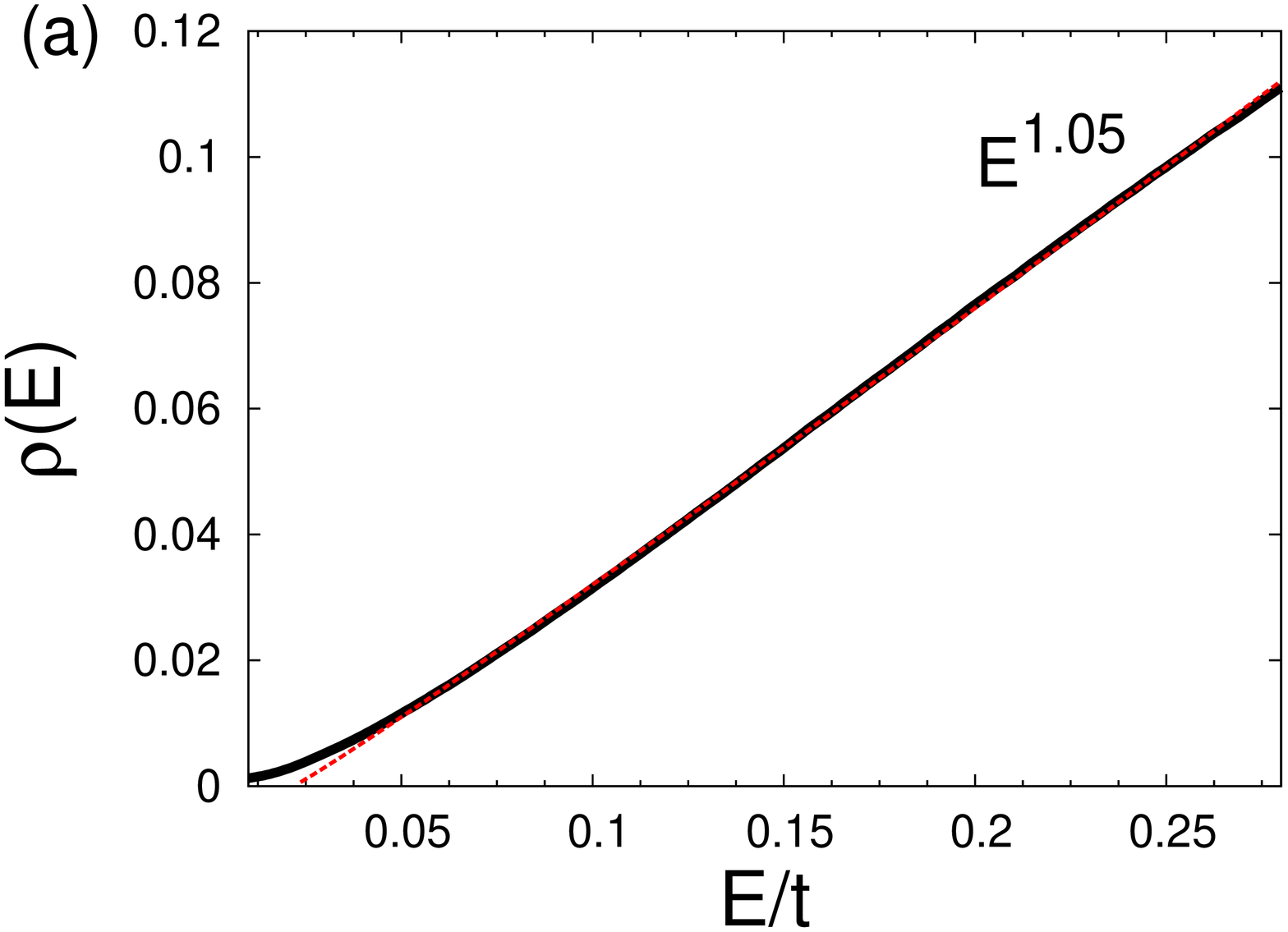}
\end{minipage}%
\begin{minipage}{.5\textwidth}
  \centering
  \includegraphics[width=0.7\linewidth,angle=-90]{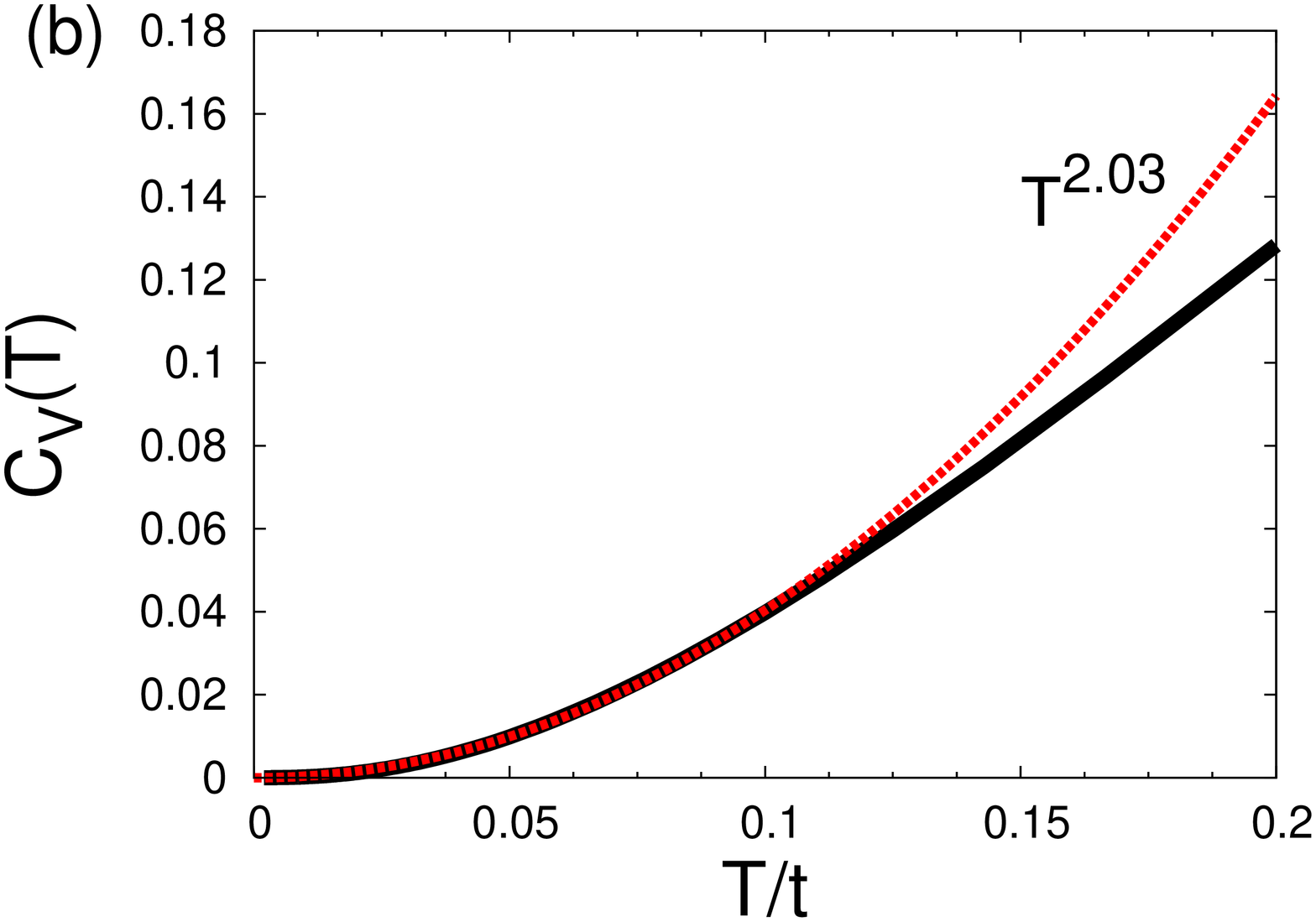}
 \end{minipage}
\caption{(Color Online) Determination of the critical exponent $z$ for potential disorder with the box distribution and $L=130$. (a) DOS as a function of energy, and (b) the specific heat as a function of temperature, both obtained at the QCP, $W_c=2.55\pm 0.05$.}
 \label{fig:3}
\end{figure*}

\section{Numerical Results}
\label{sec:numerics}
We are primarily interested in obtaining the disorder averaged DOS. As shown below
the DOS at zero energy serves as the order parameter for the SM-DM QPT. The DOS is defined as
\begin{equation}
\rho(E,L) = \frac{1}{D}\sum_{i=1}^D \delta(E-E_i),
\label{eqn:rho}
\end{equation}
where $L$ is the linear size of the system with a volume $V=L^3$, $D=4V$ is the total number of states of the model, and $E_i$ corresponds to each eigenstate.  We calculate the DOS using the numerically exact kernel polynomial method (KPM) (see Ref.~\onlinecite{Weisse-2006} for explicit details), which allows us to reach sufficiently large system sizes (also see Appendix~\ref{sec:appendix-C}). We calculate the specific heat $C_V(T) = \partial \langle E \rangle/\partial T$ by numerically integrating the density of states using the following formula for non-interacting fermions
\begin{equation}
C_V = \beta^2\int_{-\infty}^{\infty} dE \frac{\rho(E)E^2}{4\cosh(\beta E/2)^2},
\label{eqC}
\end{equation}
where $\beta = 1/T$ is the inverse temperature and we are working in units where $k_B=1$.

\subsection{Method}
Central to KPM~\cite{Weisse-2006}, we expand the density of states in terms of Chebyshev polynomials $T_n(x)$ and truncate the expansion at order $n=N_c$.  (The dependence of the numerical convergence on the parameter $N_c$ must be checked explicitly- see Appendix~\ref{sec:appendix-C} for details.)  As Chebyshev polynomials are only defined on an interval $[-1,1]$, we have to rescale the Hamiltonian to ensure that its eigenvalues fall within the corresponding range.  This can be done from a simple transformation $H = a H'+b$ where $a$ and $b$ are related to the maximum and minimum eigenvalues, which we estimate using the Lanczos method.  In short, the KPM reduces the problem of diagonalizing the Hamiltonian into calculating the coefficients of the expansion $\mu_n=\mathrm{Tr}[T_n(H')]$, which can be done using only matrix-vector multiplication, and the trace can be evaluated stochastically. For the sparse Hamiltonian matrix considered here, this can be done very efficiently, and therefore,  KPM is capable of handling system sizes much larger than what can be done by direct diagonalization for our purpose.  As is well known from the analysis of Fourier series, truncating a series expansion can lead to artificial oscillations (Gibb's oscillations) in the calculation. We filter out these spurious effects by using the Jackson Kernel~\cite{Weisse-2006}.
In Appendix~\ref{sec:appendix-C} we discuss in detail the convergence of the KPM and how to choose appropriate values for $N_c$.
For the results presented in Figs.~\ref{fig:pd} and \ref{fig:2}, we have calculated $N_c=1028$ moments such that the DOS is smooth, while considering a lattice size  $L=110$ (i.e. a system size $V=110^3$, which would be unthinkable from the exact diagonalization perspective).
As the density of states is self averaging (which we have explicitly checked), we only have to average over different realizations of the disorder to reduce the noise in the calculation and obtain a smooth DOS.
For the critical exponents, we consider $N_c=4096$ and focus on a two component model (i.e. replacing all the $\alpha_{j}$ matrices in Eq.~(\ref{eqn:ham}) with Pauli matrices $\sigma_{j}$) after isolating one of the two degenerate eigenvalues due to the axial symmetry.  In this case, we consider system sizes ranging from $L=60$ up to $L=130$, and in order to completely eliminate any statistical fluctuations we perform $100$ disorder averages.
The scaling formulas used for the analysis of the numerical data are described in Appendix~\ref{sec:appendix-A}. \\

\begin{figure*}[htbp]
\centering
\begin{minipage}{.5\textwidth}
  \centering
  \includegraphics[width=0.7\linewidth,angle=-90]{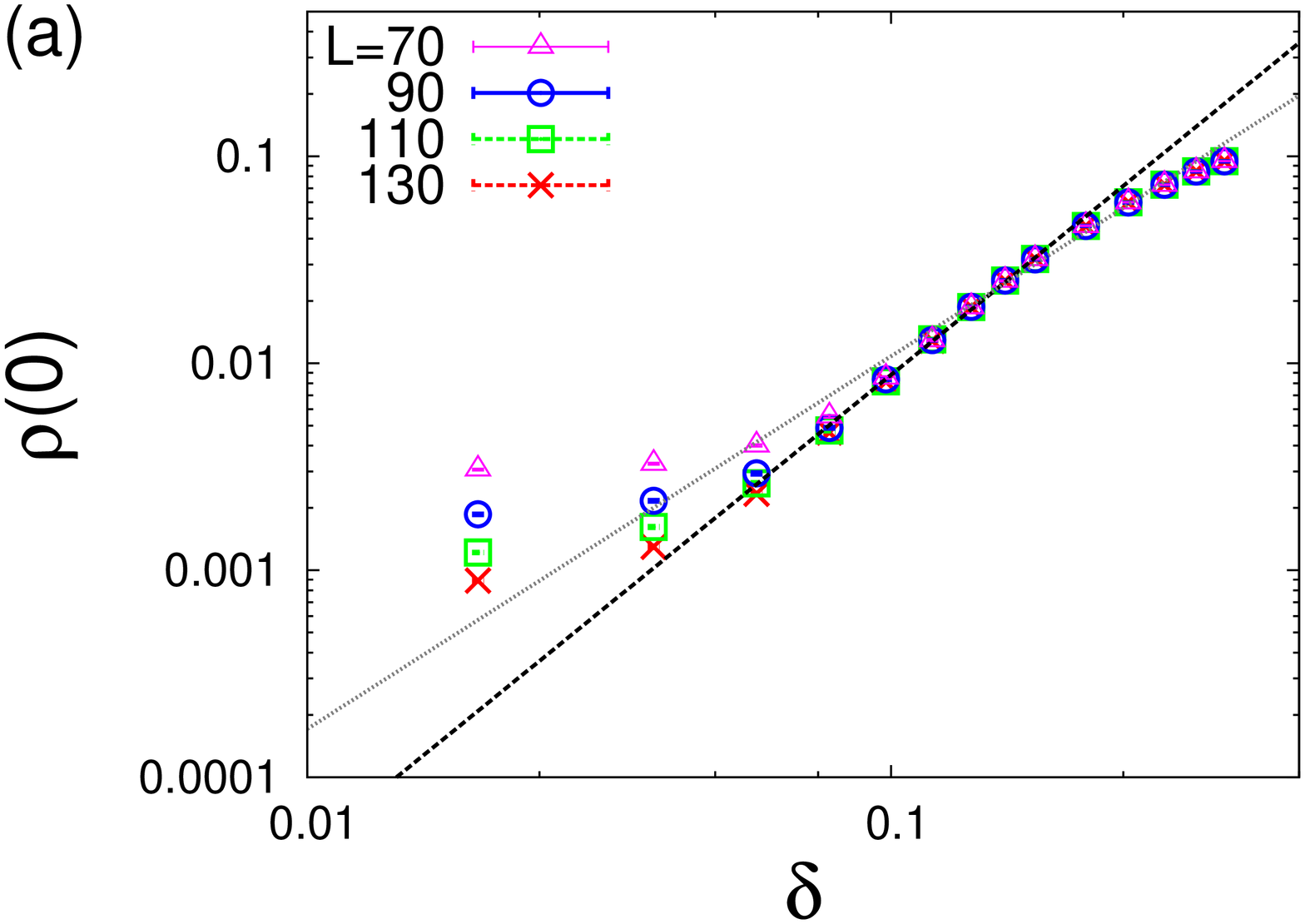}
\end{minipage}%
\begin{minipage}{.5\textwidth}
  \centering
  \includegraphics[width=0.7\linewidth,angle=-90]{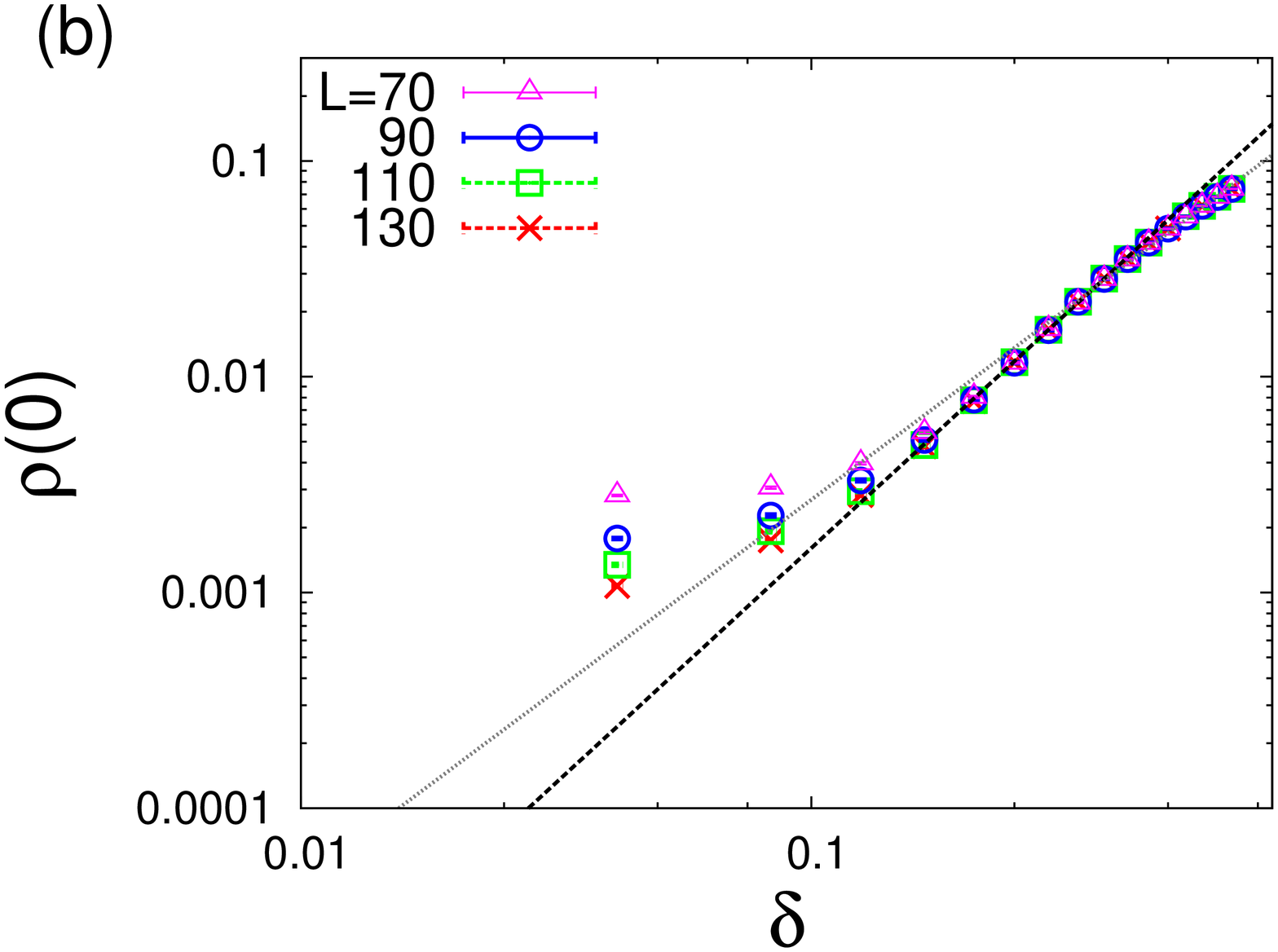}
 \end{minipage}
\begin{minipage}{.5\textwidth}
  \centering
  \includegraphics[width=0.7\linewidth,angle=-90]{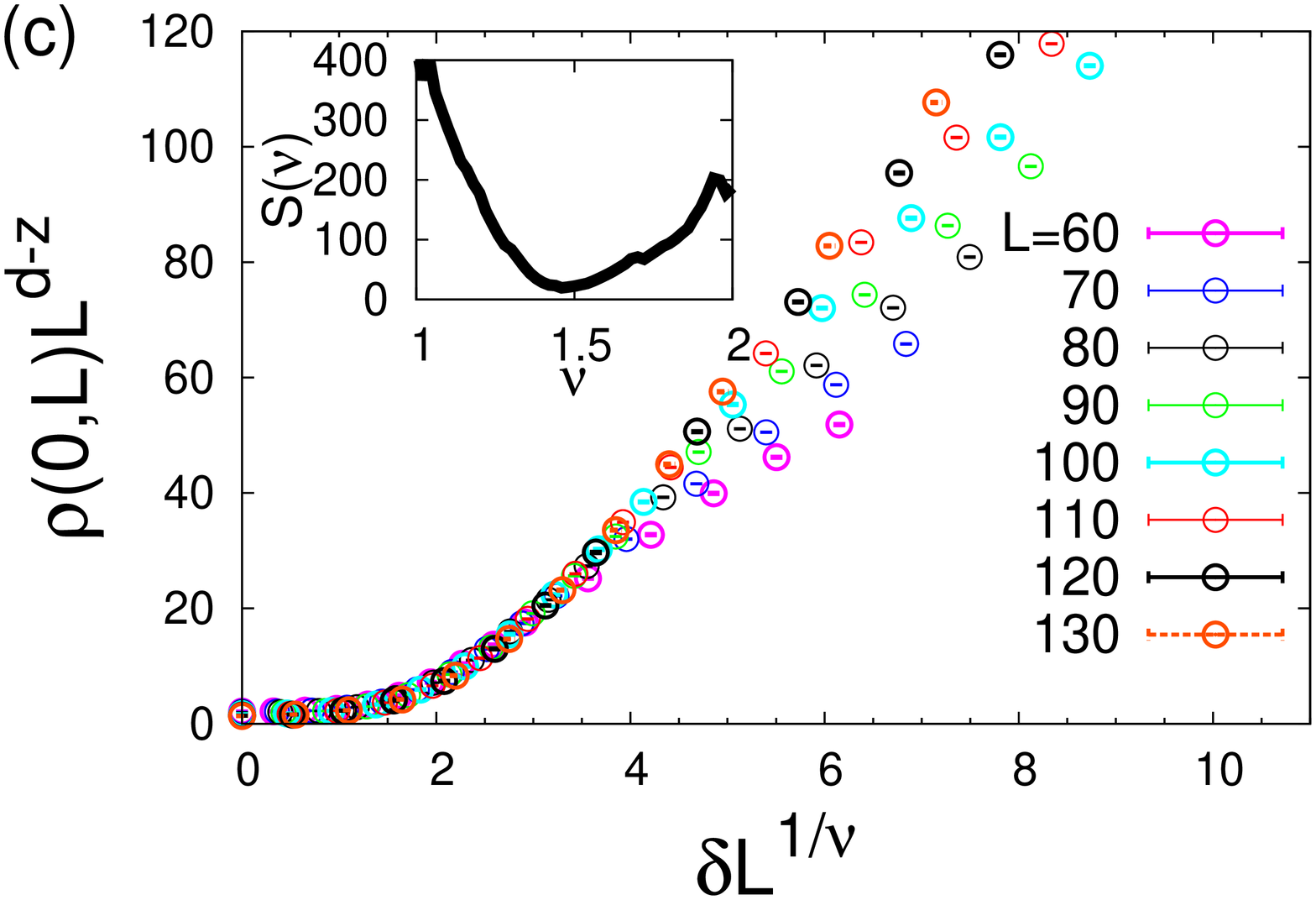}
\end{minipage}%
\begin{minipage}{.5\textwidth}
  \centering
  \includegraphics[width=0.7\linewidth,angle=-90]{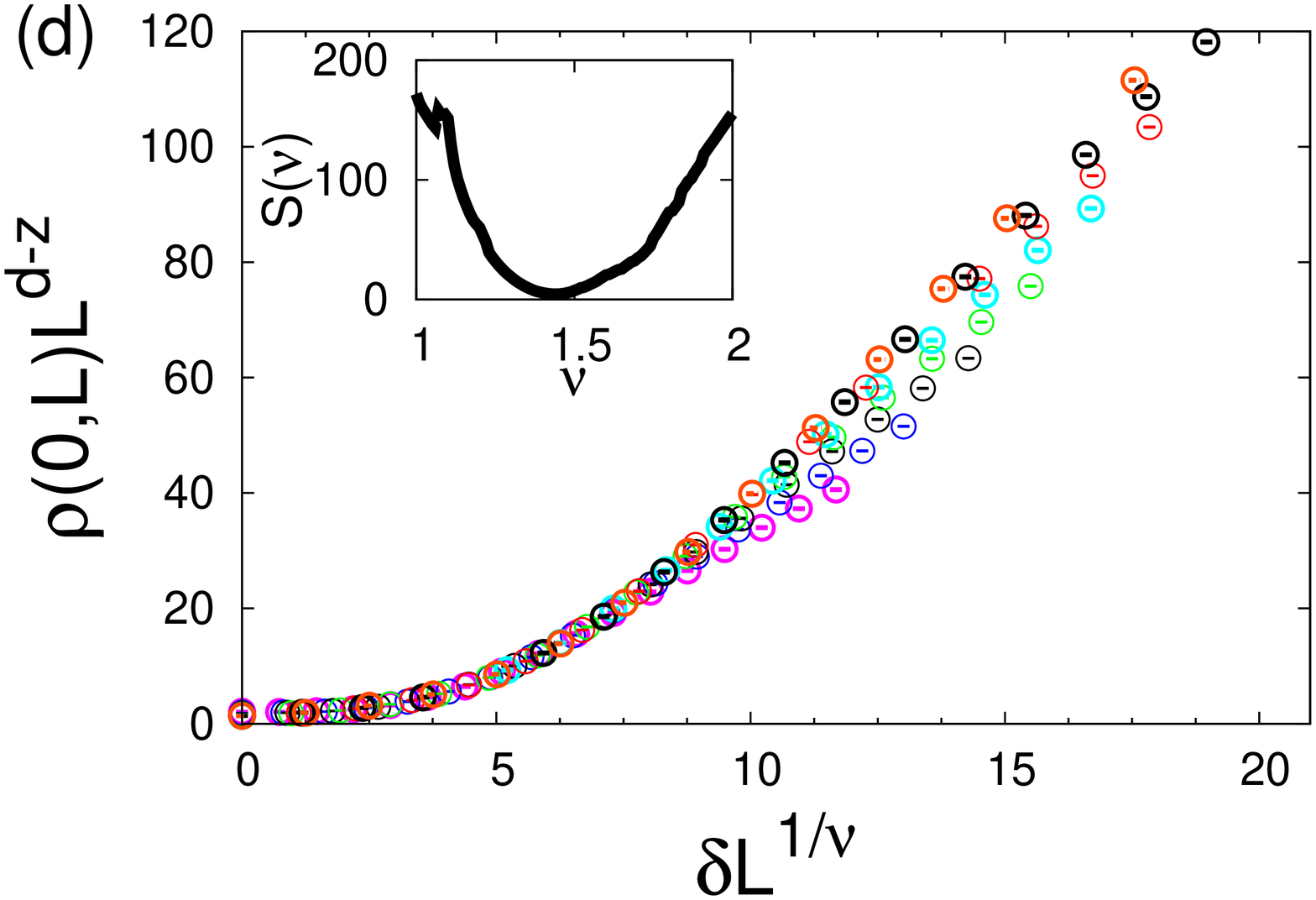}
\end{minipage}%
\caption{(Color Online) Determination of the critical exponent $\nu$ for potential disorder with the box distribution [(a) and (c)] and the Gaussian distribution [(b) and (d)]. (a) and (b) show the numerical DOS as a function of the distance to the QCP. The dashed lines are power law fits to extract $\nu$, the thick (thin) dashed lines are fits closer to (further from) the QCP.  (c) and (d) show the numerical finite size scaling and data collapse of $\rho(0,L)$, the labels for $L$ are shared between (c) and (d). (Insets) The local-linearity function~\cite{Kawashima-1993}, which yields the best scaling collapse at its minimum value. Here $\delta =|W-W_c|/W_c$ is the usual dimensionless tuning parameter defining the QPT.}
 \label{fig:4}
\end{figure*}

\subsection{Potential, Axial, and Mass Disorder}
As shown in Fig.~\ref{fig:2}(b), the DOS inside the SM phase indeed scales as $\rho(E) \propto |E|^2$, making the SM an incompressible, gapless state, which remains stable up to a critical strength of disorder. For all three types of disorder (potential, axial, and mass), $\rho(0)$ becomes finite after passing through a QCP at a finite disorder strength $W_c\approx 2.55t$, the putative SM-DM transition driven by increasing disorder. Remarkably, the microscopic critical coupling for all three types of disorder are identical within our numerical accuracy, as clearly demonstrated in Fig. 2(a).
This is a rather amazing unanticipated result since universality in critical phenomena usually refers to the universality of the critical exponents (and the scaling functions in dimensionless units), but not to the critical coupling strength (or the critical temperature $T_c$ in thermodynamic phase transitions).
The fact that the mass disorder leads to the QPT highlights the importance of intervalley scattering. At the same time the equal strength of microscopic critical couplings for all three types of disorder can not be addressed within any effective low energy theory and we have no theoretical explanation for this finding, since the critical coupling strength is not known to be a universal quantity, in contrast to critical exponents and scaling functions, in any quantum (or classical) critical phenomenon.
More work would be necessary, which is well beyond the scope of the current work, to understand this unexpected `universality in critical coupling strength' which we have numerically discovered in this problem.

 By contrast, the DOS for a two dimensional Dirac SM becomes finite for an infinitesimally weak disorder strength as shown in the inset of Fig. 2(a), which agrees with the field theoretic predictions that $d=2$ is the lower critical dimension for the disorder-tuned SM-DM QPT. The behavior of the three dimensional SM is also clearly distinct from the effects of disorder in compressible normal metals, which for an infinitesimal amount of disorder converts into a DM (from a ballistic metal with perfect conductivity). We have explicitly shown in Ref.~\onlinecite{Pixley-2015} that the DM phase (in the present model) is \emph{not} Anderson localized, and only for a large disorder strength $W_l/t \approx 8.8$ (for the box distribution) does the DM undergo Anderson localization, similar to what happens in three dimensional conventional disordered metals~\cite{Anderson-1958,Abrahams-1979,Lee-1985,Janssen-1998}. Thus, within the present calculation we find the SM (for $W<W_c$) and the DM (for $W>W_c$) phases to be stable over finite ranges of disorder in the undoped three dimensional Dirac system and we use the specific heat to determine the cross over energy scales for each phase as shown in Figs.~\ref{fig:2} (c) and (d).

\begin{figure*}[htbp]
\centering
\begin{minipage}{.5\textwidth}
  \centering
  \includegraphics[width=0.7\linewidth,angle=-90]{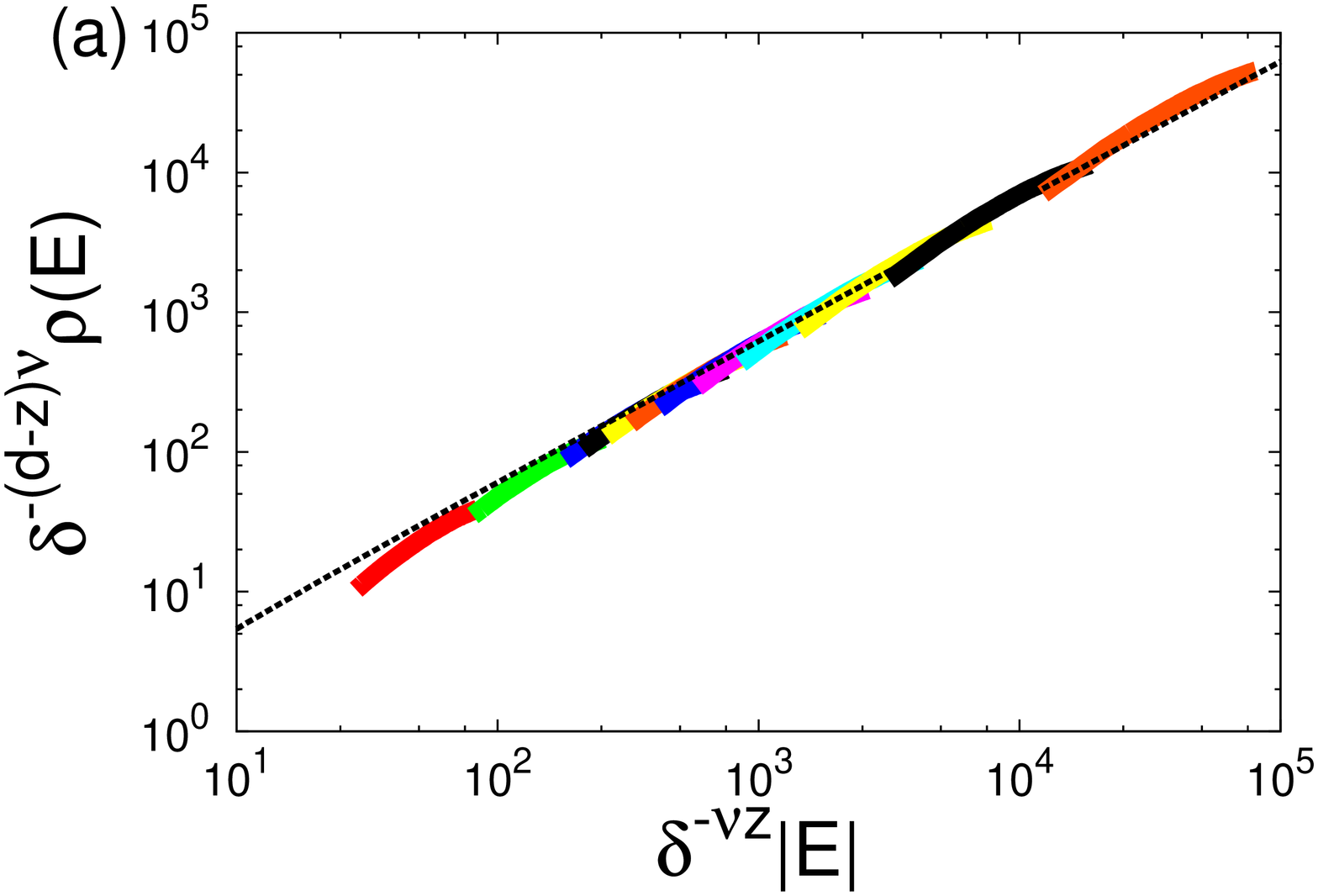}
\end{minipage}%
\begin{minipage}{.5\textwidth}
  \centering
  \includegraphics[width=0.7\linewidth,angle=-90]{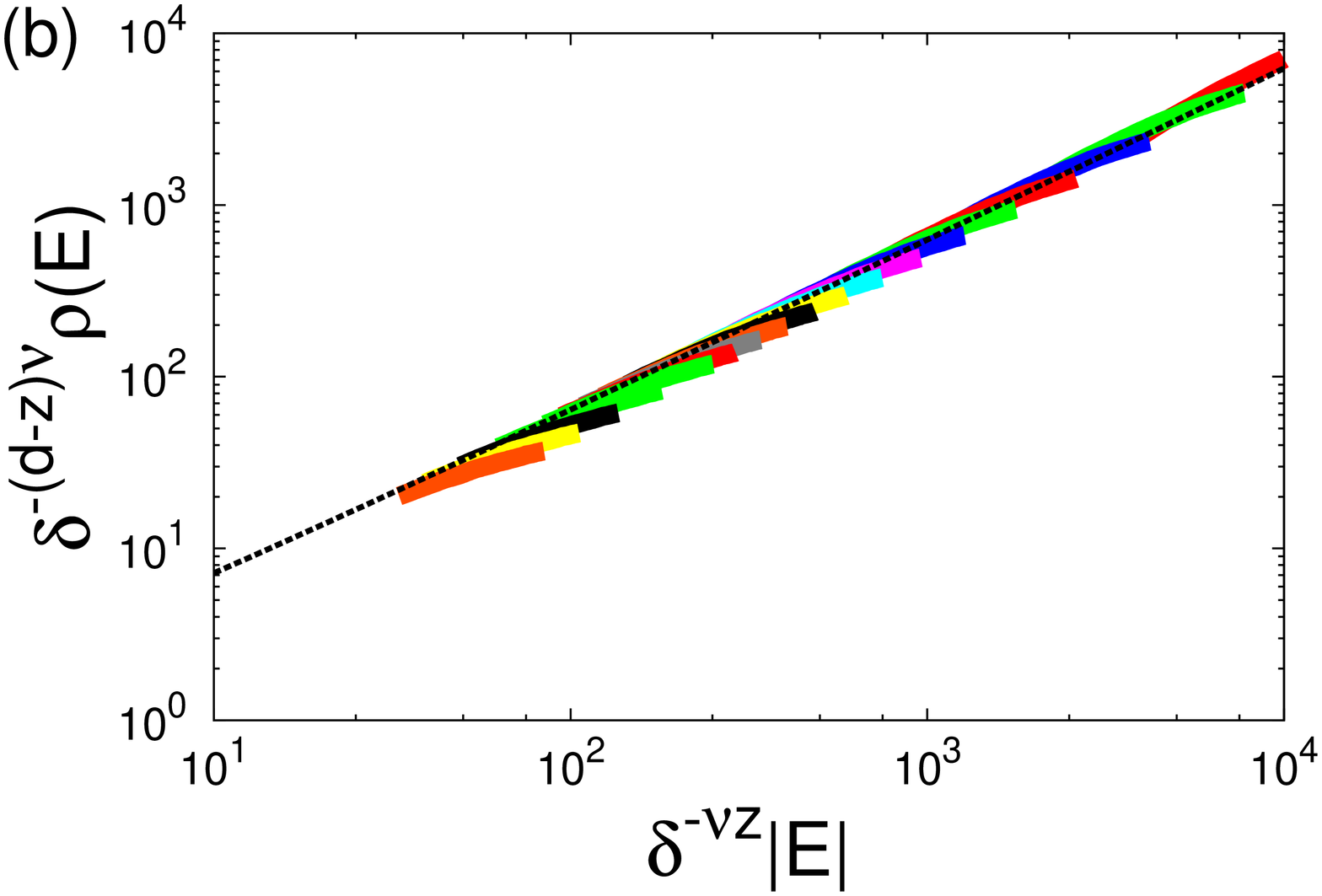}
 \end{minipage}
\begin{minipage}{.5\textwidth}
  \centering
  \includegraphics[width=0.7\linewidth,angle=-90]{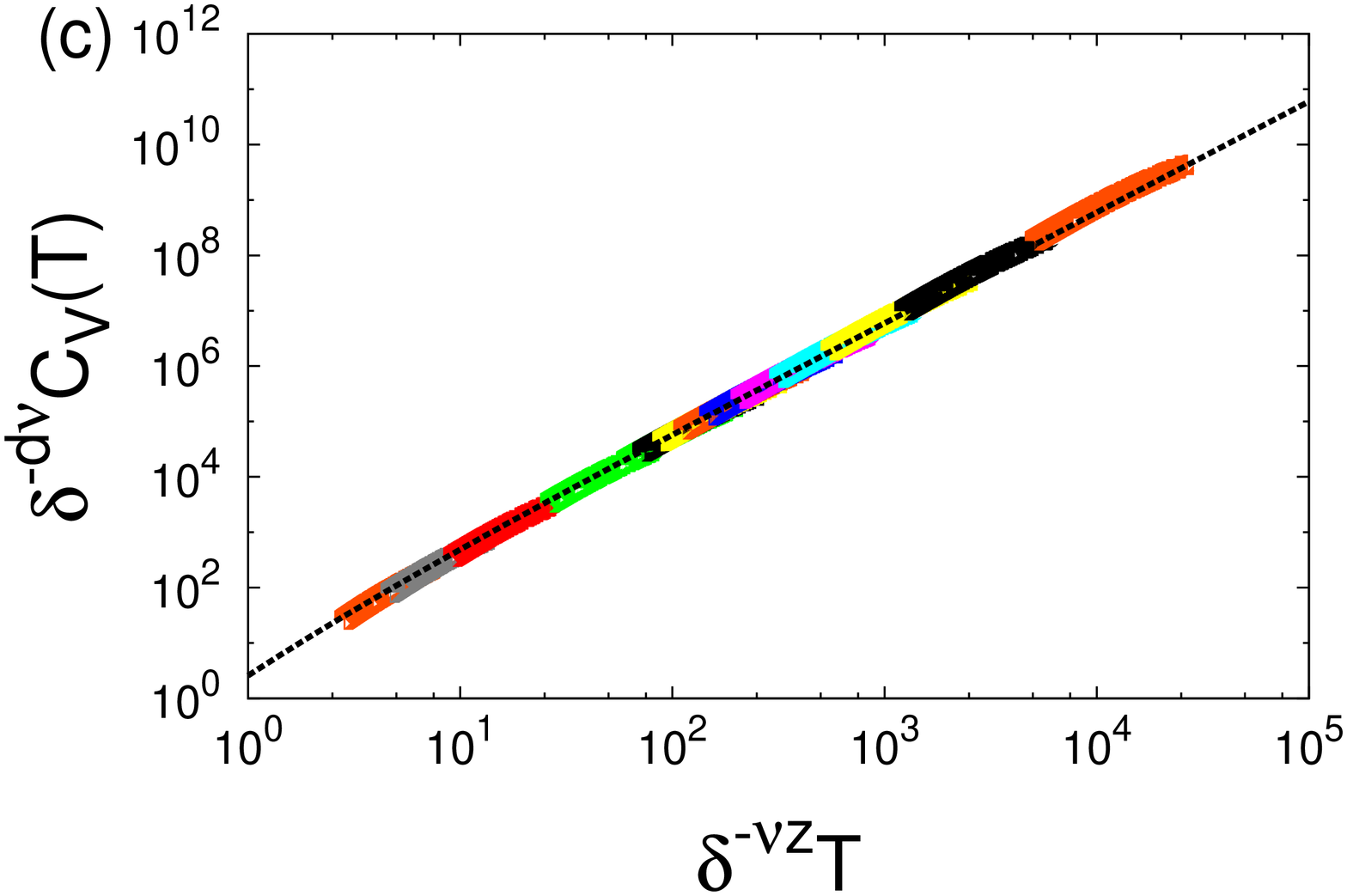}
\end{minipage}%
\begin{minipage}{.5\textwidth}
  \centering
  \includegraphics[width=0.7\linewidth,angle=-90]{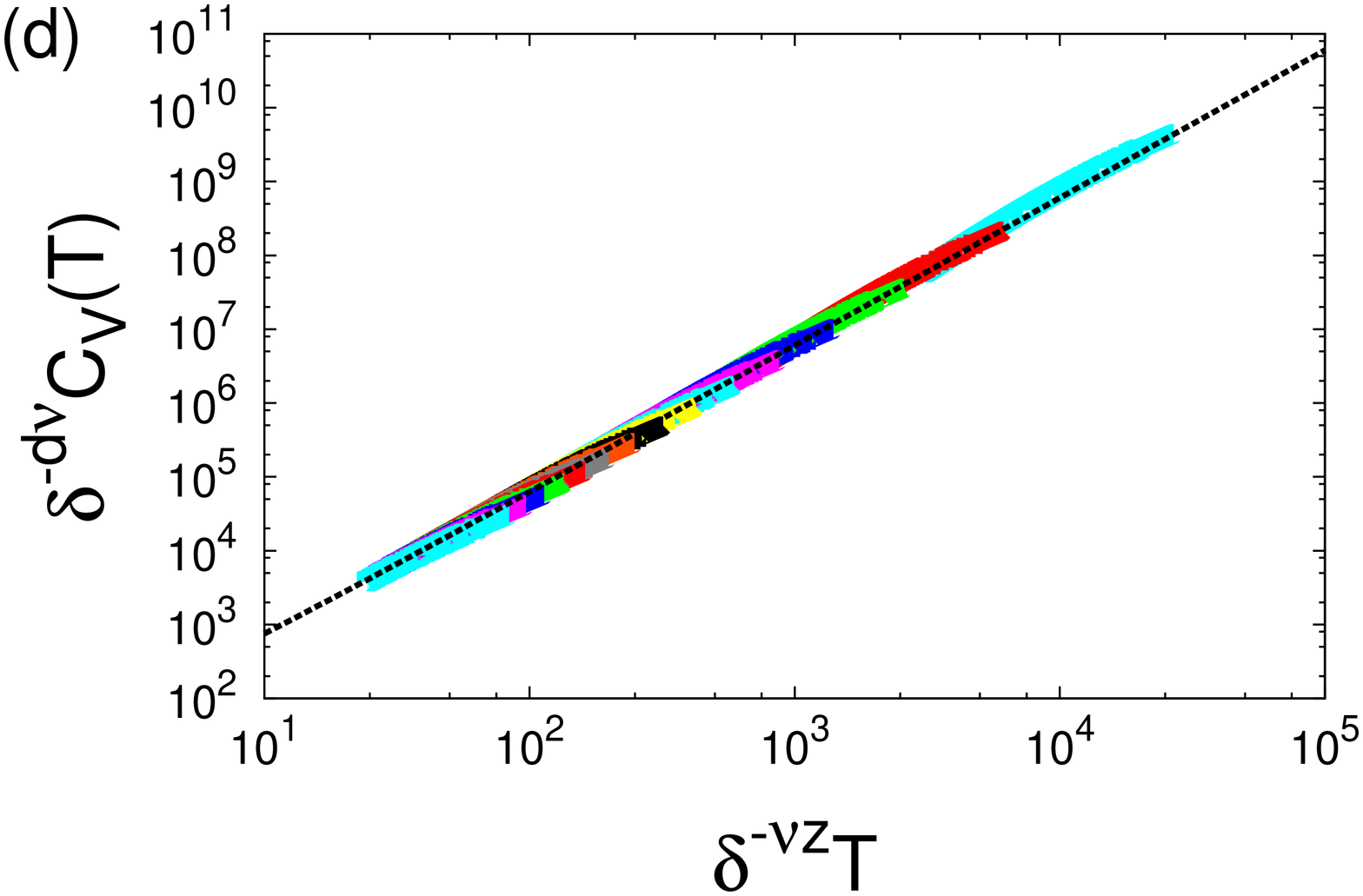}
\end{minipage}%
\caption{(Color Online) Scaling collapse of the DOS and the specific heat in the quantum critical fan $\delta^{\nu z} \ll E/t, T/t \ll \Lambda/t$ (with $\delta=|W-W_c|/W_c$) establishing that both exhibit single parameter scaling as a function of the distance to the QCP given by Eqs.~(\ref{eqn:rho_E}) and (\ref{eqn:C_T}). The dashed line is the approximate crossover functions analytically determined from the one loop beta functions (see Appendix~\ref{sec:appendix-A}); (a) and (c) are for $W<W_c$, and (b) and (d) are for $W>W_c$.  We stress that the dashed lines are not a fit, and only the non-universal amplitudes are adjusted. Despite the significant difference between the numerically obtained value for $\nu$ and the one loop result ($\nu=1$),  we find considerable agreement between the analytical and numerical crossover functions. We are using $z=1.46$ and the value of $\nu$ from the finite size scaling for the collapse of the numerical data $\nu_L\approx1.46$. These results are for the axial disorder using the box distribution, but the results for other distributions and disorder are very similar.}
  \label{fig:5}
\end{figure*}

\subsection{Universal scaling and critical exponents}
From our numerical results we now show that the density of states and specific heat obey universal scaling forms in the vicinity of the QCP, which enables us to determine the critical exponents characterizing the universality class (see Appendix~\ref{sec:appendix-A} for the details regarding the scaling ansatz).
 Assuming that the QCP is non-Gaussian (i.e. non-mean field), we apply the scaling hypothesis to the density of states and specific heat implying that they must obey the scaling forms
\begin{eqnarray}
\rho(E,L) &=&  \delta^{(d-z)\nu}\mathcal{R}(|E|\delta^{-\nu z},L^{1/\nu}\delta)
\label{eqn:rho_E}
\\
C_V(T) &=& \delta^{\nu d} \mathcal{C}(T\delta^{-\nu z},L^{1/\nu}\delta)
\label{eqn:C_T}
\end{eqnarray}
where we have introduced the reduced distance to the QCP, $\delta \equiv |W-W_c|/W_c$, and $\mathcal{R}$ and $\mathcal{C}$ are two unknown scaling functions, which are related through
\begin{eqnarray}
\mathcal{C}(T\delta^{-\nu z},L^{1/\nu}\delta)&=&\frac{T\delta^{-\nu z}}{4} \int_{-\infty}^{+ \infty} \; du \; u^2 \; \cosh^{-2} (u/2) \nonumber \\ && \times \mathcal{R}(|u| T \delta^{-\nu z},L^{1/\nu}\delta).
\end{eqnarray}

Based on the numerical calculations presented in Fig.~\ref{fig:2} (a), we have established that the QCP driven by axial or mass or potential disorder between the SM and the DM falls within the same universality class (within our numerical accuracy). Therefore, we conclude that irrespective of the underlying chiral symmetry being present or absent, the disorder driven SM-DM QPT of three-dimensional massless Dirac fermions exhibits a
manifest universality, which is one of our main results.

We determine the location of the critical point by extrapolating $\rho(0)$ to zero from the DM phase, which yields $W_c/t = 2.55 \pm 0.05$ for box disorder (and for all three different types of disorder) and $W_c/t=0.60 \pm 0.03$ for Gaussian potential disorder. Thus, rather curiously, while our numerically obtained critical disorder $W_c$ is the same for the three types of disorder we consider (i.e. potential, mass, axial), it is not for the different forms  of the disorder distribution (box or Gaussian) we use. We are able to accurately asses the disorder strength which places the model in the DM phase by considering its finite size dependence as $\rho(0)$ is $L$ independent in the DM phase, as shown in Appendix~\ref{sec:appendix-C}. It is possible that the method of extrapolation can under- or over- estimate the critical disorder strength, and we discuss below in the next subsections the implications of this in determining the critical exponents. (It turns out that the precise quantitative determination of $W_c$ is (is not) crucial for determining the critical exponent $\nu$ ($z$) accurately.)

\subsubsection{Determining $z$}
We begin by discussing the procedure we use to estimate the dynamical exponent $z$ (see Fig.~\ref{fig:3}). After determining $W_c$ by extrapolation we perform a power law fit to the energy dependence of $\rho(E)$ at $W=W_c$ [see Fig.~\ref{fig:3}(a)] and then use the scaling formula (see Appendix A)
\begin{equation}
\rho(E) \sim |E|^{d/z-1}
\end{equation}
to get $z$, here we find $z=1.46 \pm 0.05$ for both BD and GD (we only show the BD results in Fig. 3 (a)). Error bars are estimated by considering the value of $z$ due to the uncertainty of $W_c$, which is actually quite small. Even arbitrarily assuming a value $W_c$ that exceeds the error bars, for example $W_c=2.65$ (where the finite size effects are still large), yields a value of $z$ close to $1.55$ for BD, and thus an inaccurate estimate of $W_c$ does not produce a large deviation in the extracted value of $z$. We then check for consistency in the estimated $z$ by computing the specific heat, which integrates over the entire bandwidth, as shown in Fig.~\ref{fig:3}(b).  Here we use the scaling of the specific heat at the QCP
\begin{equation}
C_V(T) \sim T^{d/z}
\end{equation}
and the result from $C_V$ is $z=1.48\pm 0.05$ in good agreement with the DOS result shown in Fig. \ref{fig:3}(a).  We mention that very similar results are obtained for GD which are not shown here.  The fact that the same value of $z$ is obtained numerically from independent considerations of the DOS and specific heat gives high confidence in the numerical accuracy of our estimated dynamical exponent.

\subsubsection{Determining $\nu$}
We now turn to estimating the correlation length exponent $\nu$ (see Fig.~\ref{fig:4}). Here, we find that obtaining a precise value of $\nu$ using the KPM method alone is numerically challenging as the value is sensitive to both the accuracy of $W_c$ and the range over which the numerical results are fitted to extract the power law dependence. These features in the KPM data for the zero energy DOS has been previously pointed out in Ref.~\cite{Sbierski-2015} and therefore we analyze the data accordingly.  We first determine $\nu$ based on the scaling
\begin{equation}
\rho(0,L) \sim \delta^{\nu(d-z)},
\label{eqn:rho_d}
\end{equation}
for sufficiently large $L$ (i.e. where the data is $L$ independent) using the values of $W_c$ and $z$ we have already obtained.
We do not fit the data sufficiently close to the QCP, as there are significant finite size effects as shown in Fig.~\ref{fig:4} (a) and (b). It is well-known that the numerical data close to the QCP suffer from severe finite size effects since the correlation length exceeds the system size around the QCP.  As depicted in Fig.~\ref{fig:4} (a) and (b) using the thick dashed line, fitting a range of $\delta$ for $L=130$ starting where $\rho(0,L)$ saturates in $L$, yields a value $\nu=1.48 \pm 0.3$ for BD and $1.36 \pm 0.3 $ for GD, where the error bars are obtained by
considering the inaccuracy of $W_c$, giving consistent results with varying the range of the fit.  Fitting to a larger range of $\delta$ away from the QCP (as done in Refs.~\cite{Pixley-2015,Pixley2015disorder}) as shown in Fig.~\ref{fig:4} (a) and (b) (the thin dashed line)  yields much lower estimates $\nu=1.02 \pm 0.2$ for BD and $1.15 \pm 0.2 $ for GD.
We remark that fitting the typical DOS (i.e. the geometric averaged local DOS) data versus $\delta$ in Ref.~\cite{Pixley-2015} suffers from these same effects and the critical exponents governing the average and typical DOS at the QCP remain \emph{distinct}.

We also consider the effect of an underestimate of $W_c$ on the value of $\nu$.  Again taking the box distribution with $W_c=2.65$ for the critical disorder yields a value of $\nu=0.92$, which points to a large deviation of $\nu$ with regards to the accuracy of $W_c$.  Thus, the numerical KPM technique, while being excellent for estimating the dynamical exponent $z$, is lacking in accuracy in estimating the correlation exponent $\nu$, a situation quite common in numerics on QPTs where the numerical extraction of $\nu$ is often much more challenging as it is strongly affected by finite size and fluctuation effects.

We now come to finite size scaling and data collapse of $\rho(0,L)$.  Here we focus on the general scaling form
\begin{eqnarray}
\rho(0,L) &=&  \delta^{(d-z)\nu}\mathcal{R}(0,L^{1/\nu}\delta),
\label{eqn:rho_L}
\end{eqnarray}
in the vicinity of the QCP. Using the values we have extracted for $W_c$ and $z$, we perform data collapse for $W \ge W_c$ as shown in Fig.~\ref{fig:4} (c) and (d).  We denote the value of the correlation length exponent extracted from finite size scaling as $\nu_L$. We find excellent scaling collapse of the KPM data for $ \nu_L $ equal to $1.46 \pm 0.25$ for BD and $1.42 \pm 0.3$ for GD. Consistent with the power law fit of Eq. (\ref{eqn:rho_d}), we again find that performing the collapse over a larger range away from the QCP or accounting for an underestimate of $W_c$ yields a value of $\nu_L$ closer to $1$ (than to $1.5$).

The fact that both box and Gaussian distributions give consistent results for $\nu (\sim 1.5$ or $1$ depending on whether the scaling fit is being carried out close or away from the QCP) within the error bars provides
 evidence for the universality of the critical exponent $\nu$ (i.e. that $\nu$ is independent of the disorder distribution). However, while it is possible to estimate $\nu$ from the KPM data, it seems to suffer large fluctuations from small systematic errors (e.g. the precise value of $W_c$). Thus we make the conservative estimate for $\nu$ to lie within the range $0.9-1.5$.
Much more computationally demanding work on much larger systems would be necessary to obtain an accurate estimate of the correlation length exponent $\nu$ in this problem.
By contrast, our numerically estimated dynamical exponent $z$ is reliable and stable.

\subsubsection{Scaling in the quantum critical fan}
To establish the universal nature of the quantum critical fan
we compare the numerically determined crossover functions $\mathcal{R}$ and $\mathcal{C}$ with their approximate analytical forms (derived in Appendix~\ref{sec:appendix-A}) in Fig.~\ref{fig:5}, when $|E|>L^{-1} \delta^{-\nu z}$ and $T>L^{-1} \delta^{-\nu z}$.
Focusing on our numerical data that is restricted to lie in the non-Fermi liquid quantum critical fan (i.e. $\delta^{\nu z} \ll E/t, T/t \ll \Lambda/t$) we plot the numerical results for the DOS versus $E\delta^{-\nu z}$ and the specific heat versus $T\delta^{-\nu z}$ using the values of critical exponents $z=1.46$ and $\nu_L\approx1.46$ for the box disorder as well as the cross over functions $\mathcal{R}$ and $\mathcal{C}$ determined analytically on either side of the QCP in Fig.~\ref{fig:5}. We find that all of our data collapse onto a single curve manifesting excellent critical scaling. \emph{Even though there is a significant difference between the numerically obtained value of $\nu( \approx 1.4)$ and the approximate one loop result $(\nu=1)$, the crossover functions obtained from the two methods appear to show considerable agreement}. We stress that this is not a fit, and the only parameter that is adjusted is the non-universal coefficient of the analytical scaling functions. Finally, the consistency between our numerical calculations and the scaling hypothesis (contained in Eqs.~\ref{eqn:rho_E}) has led us to conclude that the QCP is indeed non-Gaussian and the disorder-tuned SM-DM QPT in Dirac systems cannot be described by a Gaussian theory.

Motivated by the universality of the non-Gaussian QCP, in the following section we construct the effective field theory of the long wavelength fluctuations, while accounting for the strong coupling between the underlying order parameter and the itinerant fermions.

\section{Order parameter field theory}
\label{sec:orderparameter}
Within the replica field theory formulation of disordered systems, the sample to sample fluctuations are captured in terms of a disorder induced effective statistical interaction between the fermions although the starting bare Hamiltonian is noninteracting [cf. Eqs. (\ref{eqn:ham}) and (\ref{eqn:ham2})]. The collective modes of the density fluctuations are captured by a bosonic matrix field $Q \sim \langle \psi_a \psi^\dagger_b \rangle$ and $Tr(Q)\propto \rho(0)$
acts as the order parameter. The deviation of $Tr(Q)$ from its vacuum expectation value constitutes the amplitude or the DOS fluctuations, while the transverse part of the matrix field captures the slow diffusons or Goldstone modes.
At the SM-DM QCP, both the amplitude $Tr(Q)$ and the transverse diffuson modes are gapless and remain strongly coupled with the underlying fermions.
As we show below, the average DOS controls the residue of the diffusion pole, and it therefore
vanishes as $W \rightarrow W_c^+$.
Similarly,
 the quasiparticle residue of the Dirac excitations vanishes as $W \rightarrow W_c^-$.
Consequently, inside
the critical fan the entire notion of weakly coupled quasiparticle
excitations becomes invalid and an emergent strongly coupled non-Fermi liquid
state is realized. The effective Lagrangian for the semimetal-metal QCP is therefore described by
\begin{equation}
\mathcal{L}[Q_{ab},\psi_a]=\mathcal{L}_D[\psi_a] +\mathcal{L}_B[Q_{ab}] +  i \lambda \psi^\dagger_a \psi_b Q_{ab},
\end{equation} where $\mathcal{L}_D$ describes the Dirac fermions, $\mathcal{L}_B$ is a Landau-Ginzburg functional of $Q$ up to quartic order, and
$\lambda$
is the coupling between the fermions and the bosons. This effective theory can be analyzed within a $d=4-\epsilon$ expansion around the upper critical dimension of $4$, which leads to a critical value of $\lambda \sim O(\epsilon)$. It is this strong coupling between the collective modes and the fermions which unifies the problem at hand with the underlying conceptual theme of the itinerant QCP in strange metals. 

For deriving an order parameter theory, which can capture the low energy properties of both the semimetal and the diffusive metal, we consider the product of the advanced and retarded propagators $G^R(E+\omega/2+i\eta,x)G^A(E-\omega/2-i\eta,x)$. This can capture the diffuson or Goldstone modes inside the metallic phase. For this correlation function, we can write a generating functional $e^{-S}$, where the effective action is
\begin{eqnarray}
S=\int d^dx \; \Psi^\dagger [E -\left(\frac{\omega}{2}+i\eta \right)\tau_3-H] \Psi.
\end{eqnarray} Here $\Psi^\dagger=-i(\psi^\dagger_R, \psi^\dagger_A)$, and $\tau_3=\mathrm{diag}(1,-1)$ is a Pauli matrix that operates on the retarded and advanced labels. All physical quantities such as the average DOS and the density-density correlation functions can be obtained from the generating functional after taking derivatives with respect to the appropriate source terms. For simplicity we again consider only a single Dirac cone and the case of a random scalar potential. After averaging over disorder with the replica method we arrive at
\begin{eqnarray}
S=\int d^dx \; \Psi^\dagger_a [E -\left(\frac{\omega}{2}+i\eta \right)\tau_3+ i v\alpha_j \partial_j] \Psi_a \nonumber \\ -\frac{\Delta}{2} \int d^d x \Psi^\dagger_a \Psi_a \Psi^\dagger_b \Psi_b.
\end{eqnarray}
When $\omega=\eta=0$, the retarded and the advanced sectors appear in a symmetric manner, implying there is a flavor symmetry between these two sectors. Therefore, $\eta$ acts as an infinitesimal external field for the bilinear $i \Psi^\dagger \tau_3 \Psi$, and for $\Delta>\Delta_c$ ($\rho(0)\neq 0$) this symmetry is spontaneously broken. We perform the following Hubbard-Stratonovich transformation
\begin{eqnarray}
& &\exp \left[\frac{\Delta}{2}\int d^dx \; \Psi^\dagger_a  \Psi_a \Psi^\dagger_b  \Psi_b \right] = \nonumber \\
& &\int D[Q] \exp \left[\int d^d x\left( -\frac{Tr(Q^2)}{2\Delta} +i \Psi^\dagger_a \Psi_b Q_{ab}\right)\right],
\end{eqnarray}
where $Q$ constitutes a matrix order parameter, and $\langle Q_{ab} \rangle \sim \langle \Psi_a \Psi^\dagger_b \rangle$.
By computing the susceptibilities for different ordering channels,
we find that the most favorable choice for $Q$ inside the metallic phase is $Q= i Q_0 \tau_3$, for which $\eta$ acts as the external field.
For $2<d<4$ the integral in the gap equation for $Q_0$
\begin{equation}
\frac{Q^2_0}{v^2 \Lambda^2} \int^{1}_{0} dx  \frac{x^{d-3}}{x^2+\frac{Q^2_0}{v^2\Lambda^2}}=\left (\frac{1}{\Delta_c}-\frac{1}{\Delta} \right)
\end{equation} leads to
\begin{equation}
\frac{Q_0}{v}=\Lambda \left (\frac{1}{\Delta_c}-\frac{1}{\Delta} \right)^{1/(d-2)} \sim \xi_l^{-1}.
\end{equation}
The order parameter's expectation value $Q_0$ acts as the inverse life-time $\tau^{-1}$ of the Dirac quasiparticles. Since in the mean-field calculation we have not accounted for the correction to the real part of the self-energy, which leads to a modified dispersion relation at the QCP, the mean-field solution is inadequate for capturing the scaling form $Q_0 \sim \delta^{z \nu}$. We expect that accounting for strong order parameter fluctuations will finally lead to the correct result.

After the Hubbard-Stratonovich decoupling, if we integrate out the fermion fields, the fermion bubble contributes to the $Q_{ab}-Q_{ab}$ correlation function. Since the replica indices for such a correlation function are fixed from the outset, the fermion bubble is not proportional to $N_r$ and it survives the $N_r \to 0$ limit. We can further decompose the matrix form of $Q_{ab}=Q_{ab, \mu, s}\tau_\mu \otimes \Gamma_s$, where $\Gamma_s$ are the sixteen $4 \times 4$ matrices operating on the original spinor index and $\tau_\mu$ are $2 \times 2$ matrices operating on the $R/A$ indices (we have not yet considered the Cooper channel). By computing the zero momentum part of the fermion bubble, we can show that only the conventional and chiral density channels corresponding to $\Gamma_0=I_{4 \times 4}$ and $\Gamma_5=\gamma_5$ channels are most attractive and degenerate. Therefore, only these two density channels can have gapless diffuson modes. Since the order parameter or the amplitude of $Q$ vanishes at the QCP, we can not safely integrate out the gapless fermion modes. For this reason, it is necessary to (phenomenologically) introduce a fermion-diffuson coupling (of the Yukawa form) to capture the interplay of gapless order parameter and fermionic excitations. For simplicity if we consider $E=\omega=0$, the effective order parameter field theory takes the following form
\begin{eqnarray}
\label{eqn:FT0}
\mathcal{L}[\Psi_a,Q_{ab}]&=&\mathcal{L}_D[\Psi_a]+L_B[Q_{ab}]+L_{DB}[Q_{ab},\Psi_a], \,\,\,\,\,\,\,\,\, \\
\mathcal{L}_D[\Psi_a]&=&i v\int d^dx \; \Psi^\dagger_{a}  \boldsymbol \alpha \cdot \nabla \Psi_{a}, \\
\mathcal{L}_B[Q_{ab}]&=&\int d^dx \; \bigg[ \frac{1}{2} \mathrm{Tr}(\nabla Q^\dagger \nabla Q)+ \frac{r}{2} \mathrm{Tr}(Q^\dagger Q) \nonumber \\ &&+ u_1 \mathrm{Tr} [(Q^\dagger Q)^2]+u_2 [\mathrm{Tr}(Q^\dagger Q)]^2\bigg],\\
\mathcal{L}_{DB}[Q_{ab},\Psi_a]&=&i \lambda\int d^d x \; \Psi^\dagger_a \Psi_b Q_{ab},
\label{eqn:FT}
\end{eqnarray} where the summation over repeated replica indices is implied. We are only considering $Q_{ab}$ in the regular and the axial density channels. When an external frequency is considered, it couples to $Tr[Q \tau_3]$ in the regular density channel as an external field.

At the critical point $r=0$, both regular and axial density fluctuations become gapless. If we scale $x \to x e^l$, $\Psi \to \Psi e^{(1-d)l/2}$, $Q \to Q e^{(2-d)l/2}$, we find $\lambda \to \lambda e^{(4-d)l/2}$, $u_j \to u_j e^{(4-d)l}$. Therefore, $d=4$ serves as the upper critical dimension, and the interaction effects can be addressed through a $d=4-\epsilon$ expansion, which leads to a critical point at $\lambda_c=O(\epsilon)$, $u_j=O(\epsilon)$.

In the DM phase the density-density correlation function (involving both retarded and advanced sectors) displays a diffusion pole corresponding to the gapless Goldstone modes or the diffusons, which behaves as
\begin{equation}
\Pi_d(\omega,\mathbf{k})\sim \frac{Q^2_0}{i Q_0 \omega  - \delta^{-\nu \eta_Q} \; \frac{Q^2_0}{\Lambda^2}k^2} \sim \frac{\delta^{z \nu}}{i \omega - D k^2}
\end{equation}
 where $\eta_Q$ is the anomalous dimension of the gapless collective mode. If the order parameter inside the DM phase varies as $Q_0 \sim \delta^{\beta}$, the hyperscaling relation leads to 
\begin{equation}
\beta=(d-2+\eta_Q)\frac{\nu}{2}=z\nu,
\end{equation} 
which implies $\eta_Q=2z+2-d$ (putting in our numerical estimate of $z$ yields $\eta_Q \approx 1.9$). We have also introduced the diffusion constant ($D$)
 \begin{eqnarray}
 D \sim Q_0 \delta^{-\nu \eta_Q} \sim \delta^{-\nu (z+2-d)}.
 \end{eqnarray}
 Upon approaching the QCP in $d=3$ from the metallic side, the diffusion constant diverges as $\delta^{-\nu(z-1)}$~\cite{Kobayashi-2014}. But most importantly, the residue of the diffusion pole vanishes. Therefore, the diffusion pole loses its meaning as a well defined excitation, when the QCP is approached from the metallic side. In addition, such an order parameter theory also gives rise to an anomalous dimension for the fermion field $\eta_\psi \sim O(4-d)$, which implies that the quasiparticle residue of the Dirac fermion vanishes as $\delta^{\nu\eta_{\psi} }$ when the QCP is approached from the SM side.
 
The vanishing of the quasiparticle residue of the Dirac fermions can also be addressed within the $2+\epsilon$ expansion scheme. After using the one loop RG procedure as in Ref.~\cite{Goswami-2011} the retarded propagator in the SM phase acquires the form
\begin{eqnarray}
G_R(E=0) \sim \frac{\delta^{\nu \eta_{\psi}}}{[\eta \Gamma_0 + v \delta^{(z-1) \nu}\mathbf{k}\cdot \boldsymbol \Gamma]}. \end{eqnarray}
Therefore the quasiparticle residue and the effective Fermi velocity are respectively given by $\delta^{\nu \eta_{\psi}}$ and $ v \delta^{\nu (z-1)}$. At the one loop order $\eta_\psi=(z-1)=\epsilon/2$ and $\nu=1/\epsilon$. Consequently, both the quasiparticle residue and the Fermi velocity behave as $\delta^{1/2}$. This is an artifact of the one loop analysis and at higher loop orders (beginning at $\epsilon^2$) they will follow different power laws. Hence, we conclude that the quasiparticle residue and the effective Fermi velocity of the Dirac fermion vanish, as we approach the QCP from the semimetal side. 
 
Consequently, the residue of the two different types of quasiparticles vanish while approaching the QCP from either side.  Thus, the critical region of this system can not have any simple quasiparticle description, qualitatively similar to what is envisaged for QC regions in correlated materials.
 Thus, the orange region in our numerically obtained quantum phase diagram of Fig.~\ref{fig:pd} is a finite-temperature `non-Fermi liquid' quantum critical crossover regime arising from the critical fluctuations of the non-Gaussian QCP underlying the SM-DM QPT in the three dimensional Dirac materials.

To summarize this section, the finite expectation value of the DOS [as shown numerically in Fig.~\ref{fig:2}(a)] on the metallic side and its critical fluctuations in the vicinity of the QCP can be described by an order parameter in the form of the lagrangian $\mathcal{L}_B$ for the matrix field $Q$. While approaching the QCP from the SM phase, the quasiparticle residue of the underlying Dirac fermions (as described by $\mathcal{L}_D$) vanishes continuously as $\delta \rightarrow 0$. Thus, the gapless fermionic and bosonic degrees of freedom must be treated on an equal footing and this has led us to construct the effective action in Eqs. (\ref{eqn:FT0})-(\ref{eqn:FT}), which governs all the long wavelength degrees of freedom. This effective theory retains both longitudinal and transverse fluctuations of the $Q$ matrix, as required by the vanishing of the average DOS at the QCP. This should be contrasted with the familiar nonlinear sigma model description of the Anderson localization at stronger disorder~\cite{Pixley-2015}, where the average DOS remains noncritical across the transition. The nonlinear sigma model is obtained by integrating out the gapped ballistic fermions and the gapped longitudinal fluctuations of the $Q$ matrix inside the metallic phase (below the energy scale of $\tau^{-1} \sim Q_0$). 

 In our earlier work~\cite{Pixley-2015} on the phase diagram of a disordered DSM, we have noticed a significant difference between the critical exponents for the typical DOS at the SM-DM QCP and the QCP describing an Anderson localization transition (at stronger disorder). We believe that the differences between the multifractal properties at these two QCPs are inherited from (i) the presence or absence of itinerant, ballistic fermions, and (ii) the presence or absence of gapless longitudinal mode of the $Q$ matrix. A detailed analysis of different correlation functions and the multifractal properties associated with the field theory in Eqs. (\ref{eqn:FT0})-(\ref{eqn:FT}) is beyond the scope of this work and remains an open problem for the future. The Dirac SM and the DM phases preserve time reversal and inversion symmetries, and do not display any topological transport properties such as anomalous Hall effect. For this reason, there is no Chern Simons term in the effective field theory. By contrast, a time reversal symmetry breaking Weyl semimetal phase and the resulting diffusive metal possess an anomalous Hall effect. In the effective field theory this arises due to a topological Chern Simons term~\cite{Altland-2015}.

One direct implication of our effective field theory is the existence of an upper critical dimension of $d_u=4$ for the disorder-tuned SM-DM transition in Dirac-Weyl systems (in addition to the known lower critical dimensionality of $d_l=2$) . For dimensions $d \geq d_u$, the correlation length exponent $\nu$ should be given by the mean-field value $\nu=1/2$ and hyperscaling relations will be violated. It will be interesting to check this observation in future numerical studies on higher dimensional models of a Dirac SM. Finally, the existence of an upper critical dimension for the SM to DM QCP will further distinguish this transition from Anderson localization, which does not possess an upper critical dimension. 

We mention here that we do not, however, have an explanation for our numerical finding of the exact $z (\approx 1.5)$ being equal within the numerical accuracy to the one-loop theoretical value ($z=d/2$).  Whether this is a mere coincidence or a deeper truth (i.e. somehow all higher-loop corrections cancel out) is unknown at this stage.  Of course, our numerical error in estimating $z$, while being small, is not zero, and thus small corrections in the exact theoretical value of $z$ (well less than $10\%$) away from $z=1.5$ cannot be ruled out without more numerical work with much larger systems.  There are well-known examples in the literature for numerical and one-loop theoretical $z$ values being very close to each other purely fortuitously in completely different contexts of dynamical phase transitions~\cite{Lai-1991,Kim-1994,Janssen-1997}. 

\section{Discussion and Conclusion}
\label{sec:conclusions}
The experimentally observed non-Fermi liquid scaling behavior of many correlated metals~\cite{Hewson-book,Ishiguro-1990,Sachdev-2003,Stewart-2011} over a significant range of temperatures is usually reconciled with the existence of a large quantum critical fan driven by the putative QCP, whose intrinsic properties are often not well-understood. It is generally believed that the generic behavior of the `strange metal' within the finite-temperature critical fan regime is non-Fermi liquid like because of strong quantum fluctuations arising from the QCP (which may often be `hidden' or `inaccessible' experimentally), but no general proof exists.
For describing such a metallic or itinerant QCP, it is necessary to address both the gapless fermionic degrees of freedom and the bosonic order parameter fluctuations on an equal footing, which is a highly challenging task for analytical calculations, particularly for strongly correlated systems, where even the minimal model is intractable with or without the QCP. In this work even though we have focused on non-interacting massless Dirac fermions we have been able to make various connections to the general notion of scaling phenomena of itinerant QCPs. First, we find the appearance of a broad quantum critical fan at finite temperatures. Second, we have succeeded in constructing an effective field theory of bosonic fluctuations (captured here by the $Q$ matrix for diffusive modes) that are strongly coupled to itinerant electron degrees of freedom (i.e. massless Dirac fermions). Third, we have shown how the general notion of quasiparticle excitations breaks down (via their residue) in the quantum critical fan region even for our noninteracting model, explicitly demonstrating that a QCP by itself, independent of the underlying model being interacting or not, leads to a generic non-Fermi liquid behavior. However, as we are dealing with non-interacting fermions, it is important to contrast the version of itinerant quantum criticality we obtain here with the form that is applicable to strongly correlated electronic systems. We are dealing with free fermions and the interaction is statistical and only due to disorder (only elastic scattering). Consequently, the effective field theory will not be a dynamical one (with inelastic scattering) as we have found. This is in sharp contrast to that of the strongly correlated systems where the dynamics of the bosonic field is also important and must be incorporated in the theory on an equal footing with the fermionic field.

Through a precise numerically exact calculation using the KPM technique and an approximate field theoretical analysis, we have established that the
three-dimensional Dirac SM phase undergoes a disorder-tuned QPT into a DM state for different types of disorder (potential, axial, and mass), which belong to the same universality class (and surprisingly, even with the identical critical coupling for different disorder types) of an itinerant QCP.
We find the critical exponents for box and Gaussian distributions agree to within numerical accuracy (with $z=1.46 \pm 0.05$ and $\nu \in[0.9,1.5]$), which points to the universal nature of the QCP. For this
model of an itinerant QCP, the universal scaling functions obtained from numerical and analytical methods show a remarkable agreement, and thus can serve as a conceptual framework for addressing other itinerant critical phenomena in diverse correlated metallic systems. Experimentally, a dilute three-dimensional Dirac material such as Na$_3$Bi with a small value of the Fermi energy $E_F \ll \Lambda$ associated with residual doping, can be a promising material for verifying our theoretical results.
In a real system, the residual unintentional doping always produces a finite Fermi energy shifting the system from the zero chemical potential Dirac point, but as long as this doping-induced Fermi energy is lower than the temperature, one expects intrinsic Dirac point behavior to prevail, leading to the manifestation of the QPT studied in this work.  One experiment could be the measurement of the specific heat in the quantum critical regime which, according to our theory, should manifest a non-Fermi liquid like quadratic temperature dependence associated with the critical quantum fluctuations.  One key aspect of this QCP is its fundamental non-Gaussian nature.
Our KPM-based estimate of the correlation length exponent $\nu$, however, has rather large error bars, and much more work would be necessary (well beyond the scope of the current work) to obtain a precisely numerically accurate value of $\nu$ in the SM-DM QPT in three dimensions.

Finally, we comment on the possible role that `Griffiths physics' (associated with rare fluctuations) might play in the SM-DM QPT studied in the current work.  The physics of rare regions (or the so-called ``Griffiths physics'') is not captured by the long wavelength field theory developed in the present manuscript.
Recently it has been suggested that rare regions affecting the SM-DM QCP can induce a finite DOS at the
Dirac point for an infinitesimally weak disorder~\cite{Nandkishore-2014},
thus converting the SM phase  into effectively a DM phase. Compared to the box distribution,
the unbounded tails of the Gaussian disorder distribution are expected to enhance these rare fluctuations, but we get the same universal QCP properties for both box and Gaussian distributions without seeing any obvious effects of rare regions.
 Within our numerical calculations
we have not found any evidence of rare region effects, as the results for both distributions agree
with the field theoretical predictions for the existence of the transition (e.g. considering the problem in two versus three dimensions).
 However, the detection of rare region effects can be quite subtle (due to a possibly exponentially small DOS contribution from the rare regions) and therefore the possible existence of rare regions still
  cannot be ruled out completely (e.g. the regions are rarer than our numerical system
  size or the rare regions being independent of the QCP itself or the exponentially small DOS introduced by the rare regions are simply too small to show up in our KPM simulations), which necessitates a detailed focused numerical study on its own.
  Although we cannot rule out the possibility of rare regions based on our study, we can assert that the disorder-tuned SM-DM QPT is well-captured by our theory and numerics once any contribution from rare fluctuations are subtracted out.  It is important in this context to emphasize that the real Dirac systems are likely to have random impurity-induced `puddles' or spatial density inhomogeneities (i.e. random spatially  nonuniform doping) around zero energy (i.e. the Dirac point) any way, as is ubiquitous in graphene~\cite{DasSarma-2011}, leading to a locally fluctuating chemical potential masking the QCP in a way similar to what rare regions would do.  Any experimental study of the QPT itself must find a reasonable way of subtracting out these additional effects
  (from rare regions and/or puddles)
  in order to capture the underlying critical behavior.

\section*{Acknowledgements} We have benefited from insightful discussions with Sudip Chakravarty, Qimiao Si, Steven Kivelson, Subir Sachdev, David Huse,  A. Peter Young, Matthew Foster, Bj{\" o}rn Sbierski, and Piet Brouwer. This work is supported by JQI-NSF-PFC, LPS-MPO-CMTC, and Microsoft Q. The authors acknowledge the University of Maryland supercomputing resources (http://www.it.umd.edu/hpcc) made available in conducting the research reported in this paper.

\appendix

\section{Scaling formulas} \label{sec:appendix-A}
 In the vicinity of the QCP, we have two divergent scales, namely the spatial correlation length $\xi_l \propto \delta^{-\nu}$ and the temporal correlation length $\xi_t \sim \xi^z_l \propto \delta^{-\nu z}$. The universal scaling functions are determined by these two divergent scales. Quite generally for an observable $\mathcal{O}$ we have the scaling formula,
\begin{equation}
\mathcal{O}(L, E, T) \sim \xi^{\Delta_{\mathcal{O}}}_{l}F_{\mathcal{O}}(L\xi^{-1}_{l}, E \xi^{z}_{l}, T \xi^{z}_{l}),
\end{equation} where $L$, $E$ and $T$ respectively denote the system size, the energy and the temperature, and $\Delta_{\mathcal{O}}$ is the scaling dimension of $\mathcal{O}$. For example, the scaling dimensions $\Delta_{\mathcal{O}}$ for the average density of states, the free energy density, the specific heat and the longitudinal conductivity are respectively given by $-(d-z)$, $-(d+z)$, $-d$, and $-(d-2)$. The leading scaling behavior is determined by $\mathrm{max}\{L^{-1},E,T,\xi^{-1}_{l} \}$. Since there are two fixed points respectively with $z=1$ and $z=3/2$ (at one loop), the crossover scaling functions are quite nontrivial.

For example, first consider the average density of states
\begin{equation}
\rho(L,E) \sim \xi^{-(d-z)}_{l}F_{\mathcal{\rho}}(L\xi^{-1}_{l}, E \xi^z_l) .
\end{equation}
At zero energy this becomes
\begin{equation}
\rho(L,0) \sim \delta^{\nu(d-z)} f(L\delta^{-\nu})\sim L^{-(d-z)}g(L^{1/\nu}\delta).
\end{equation} When $L \ll \xi_l \sim \delta^{-\nu}$, we are in the quantum critical regime and
\begin{equation}
\rho(L,0) \sim \frac{1}{L^{d-z}},
\end{equation} reflects the scale invariance. By contrast, inside the metallic phase ($\Delta >\Delta_c$)
\begin{equation}
\rho(L,0) \sim \delta^{\nu(d-z)}.
\end{equation} On the semimetal side we rather expect
\begin{equation}
\rho(L,0) \sim L^{-(d-1)}.
\end{equation} These finite size scaling properties constitute the basis for the analysis of our numerical results.
In addition, we consider a large enough system size, $L \gg (\xi_l, v/E)$. In this case, the scaling behavior is determined by the interplay between $v/E$ and $\xi_l^z$. The scaling function now behaves as
\begin{equation}
\rho(E) \sim \delta^{\nu (d-z)} h(E \delta^{-\nu z}).
\end{equation} When, $E \gg \delta^{\nu z}$, we are in the quantum critical regime and obtain the following scale invariant answer
\begin{equation}
\rho(E) \sim |E|^{d/z-1}.
\end{equation} Inside the metallic phase ($\Delta > \Delta_c$), the leading behavior is given by
\begin{equation}
\rho(E) \sim \delta^{\nu(d-z)}.
\end{equation} By contrast, inside the Dirac semimetal phase we obtain
\begin{eqnarray}
\rho(E) \sim c(\delta) E^{d-1} \sim \delta^{-d(z-1)\nu} |E|^{d-1} .
\end{eqnarray} 

We can gain considerable analytical insight regarding the crossover functions by integrating the one-loop renormalization group equations. 
The one loop RG equations for potential and axial disorder are~\cite{Goswami-2011}
\begin{equation}
z(l)=1+\Delta, \: \frac{d\Delta}{dl}=-(d-2) \Delta + 2 \Delta^2.
\end{equation} 
We simultaneously solve the beta function for the disorder coupling $\Delta$ and
\begin{equation}
\frac{d \epsilon}{dl}=z(l) \Delta=(1+\Delta)\epsilon,
\end{equation} where $\epsilon=E/(v\Lambda)$ is the dimensionless quantity defined from the energy $E$. The dimensionless temperature $t=k_BT/(v\Lambda)$ also satisfies the same flow equation as $\epsilon$. From these solutions we can identify two RG flow times associated with two divergent scales. The disorder coupling $\Delta(l)$ is given by
\begin{equation}
\Delta(l)=\frac{\Delta_c}{1+\left(\frac{\Delta_c}{\Delta_0}-1\right)e^{(d-2)l}},
\end{equation} where $\Delta_0=\Delta(l=0)$ is the bare value of the coupling constant. On the SM side $\Delta_0<\Delta_c$, the renormalized disorder coupling monotonically decreases and satisfies the asymptotic form
\begin{equation}
\Delta(l) \approx \frac{\Delta_0}{\delta} e^{-(d-2)l}.
\end{equation}
In contrast, for the metallic side $\Delta_0 >\Delta_c$ the renormalized disorder coupling diverges at the RG scale
\begin{equation}
e^{l_1}=\xi_l \Lambda \sim |\delta|^{-\nu},
\end{equation} which defines the correlation length $\xi_l$ and the scaling exponent $\nu=1/(d-2)$. By contrast, the divergent time scale has to be found from the solution of the differential equation for energy
\begin{eqnarray}
\epsilon(l)=\epsilon(0)\left(\frac{\Delta(l)-\Delta_c}{\Delta(0)-\Delta_c}\right)^{z \nu}\left(\frac{\Delta_0}{\Delta(l)}\right)^\nu \nonumber \\
=\left(\frac{\Delta_c}{\Delta_0}\right)^{(z-1)\nu}\frac{\epsilon(0) e^{z l}}{\left[1+\left(\frac{\Delta_c}{\Delta_0}-1\right)e^{l/\nu}\right]^{(z-1)\nu}}.
\end{eqnarray} For that we need to find the RG scale $l_2$ as a function of the bare energy $\epsilon(0)$ and $\delta$ by imposing the condition $\epsilon(l_2) \sim 1$. This results into the following equation for $l_2$
\begin{equation}
\left(\frac{E}{v\Lambda}\right) \; e^{z l_2} \left(\frac{\Delta_c}{\Delta}\right)^{(z-1)\nu}=\left[1+\left(\frac{\Delta_c}{\Delta}-1\right)e^{l_2/\nu}\right]^{(z-1)\nu}.
\end{equation} When $\Delta=\Delta_c$, $l_1 \to \infty$, and the infrared cutoff is solely determined by $l_2$ where
$$e^{l_2}\sim |E|^{-1/z}.$$ Inside the semimetal phase $\Delta \to 0$, and again the infrared cutoff is determined by $l_2$ with $e^{l_2} \sim E^{-1} \delta^{(z-1)\nu}$. Since $\rho$ has the scaling dimension $-(d-z)$, we find $\rho \sim |E|^{(d/z-1)}$ inside the critical fan, and $\rho \sim c(\delta) |E|^{(d-1)}$ inside the semimetal phase, with the $c(\delta)$ quoted in the previous paragraph.

The explicit solution for $e^{l_2}$ in $d=3$ inside the quantum critical fan is obtained by finding the roots of a cubic equation. When $\Delta_0>\Delta_c$, we have the following cubic equation
\begin{equation}
e^{3l_2}+\frac{\delta}{\epsilon^2(0)}e^{l_2}-\frac{1+\delta}{\epsilon^2(0)}=0,
\end{equation} which has the only one real root given by
\begin{eqnarray}
e^{l_2}&=&2\sqrt{\frac{\delta}{3\epsilon^2(0)}}\sinh \left[\frac{1}{3}\sinh^{-1} \left \{ \frac{(1+\delta)3\sqrt{3}\epsilon(0)}{2\delta^{3/2}} \right \} \right] \nonumber \\
& \approx & \frac{3 \xi_l \Lambda}{x} \sinh \left(\frac{1}{3}\sinh^{-1} x \right), \: \: \mathrm{for} \: \delta <<1,
\end{eqnarray} with the anticipated dimensionless variable $x= \frac{3\sqrt{3}\epsilon(0)\delta^{-3/2}}{2}$. When $x>>1$ we have the quantum critical behavior, which eventually gives away to the metallic behavior for $x<1$ when the infrared scale is determined by $e^{l_1}$ or the correlation length. When $\Delta_0<\Delta_c$, the cubic equation becomes
\begin{equation}
e^{3l_2}-\frac{\delta}{\epsilon^2(0)}e^{l_2}-\frac{1-\delta}{\epsilon^2(0)}=0.
\end{equation} When $ \frac{(1-\delta)3\sqrt{3}\epsilon(0)}{2\delta^{3/2}}>1$ the only real root is given by
\begin{eqnarray}
e^{l_2}&=&2\sqrt{\frac{\delta}{3\epsilon^2(0)}}\cosh \left[\frac{1}{3}\cosh^{-1} \left \{ \frac{(1-\delta)3\sqrt{3}\epsilon(0)}{2\delta^{3/2}} \right \} \right] \nonumber \\
& \approx & \frac{3 \xi_l \Lambda}{x} \cosh \left(\frac{1}{3}\cosh^{-1} x \right), \: \: \mathrm{for} \: \delta <<1.
\end{eqnarray} For $ \frac{(1-\delta)3\sqrt{3}\epsilon(0)}{2\delta^{3/2}}<1$, there are three real roots, and only one of them is positive. This root is given by
\begin{eqnarray}
e^{l_2}&=&2\sqrt{\frac{\delta}{3\epsilon^2(0)}}\cos \left[\frac{1}{3}\cos^{-1} \left \{ \frac{(1-\delta)3\sqrt{3}\epsilon(0)}{2\delta^{3/2}} \right \} \right] \nonumber \\
& \approx & \frac{3 \xi_l \Lambda}{x} \cos \left(\frac{1}{3}\cos^{-1} x \right), \: \: \mathrm{for} \: \delta <<1,
\end{eqnarray} where the inverse trigonometric function is restricted to the first quadrant. For small $\delta$, $\delta e^{l_2}$ is again a scaling function of $\epsilon(0) \delta^{-3/2}$. For energies satisfying $\epsilon(0) \delta^{-3/2}>> 1$ or $x>>1$ inside the quantum critical fan $\cosh$ and $\sinh$ terms determine the $z=3/2$ quantum critical behavior, while inside the SM phase the $\cos$ term leads to the $z=1$ critical behavior. Therefore, the crossover from the $z=3/2$ to $z=1$ scaling is appropriately captured by how the $\cosh$ term transforms into a $\cos$ term.

In the thermodynamic limit the total number of states per unit volume can be estimated to be
\begin{equation}
\frac{N(\epsilon)}{L^3} \sim \Lambda^d e^{-d l_2} = \xi^{-d}_{l} f^{-d}(x).
\end{equation} Since the average DOS can be obtained by differentiating $N(\epsilon)L^{-d}$ with respect to $\epsilon$, we can write
\begin{equation}
\rho(\epsilon) \sim \frac{\Lambda^2 \delta^{3/2}}{2 \sqrt{3} v} g(x),
\end{equation} where the crossover scaling function $g(x)$ is determined by
\begin{eqnarray}
g(x)&=&\frac{x^2}{\sinh^3 \left(\frac{1}{3} \sinh^{-1} x \right)} \left[1-\frac{x \coth \left(\frac{1}{3} \sinh^{-1} x \right)}{3 \sqrt{1+x^2}}\right], \nonumber \\
& & \mathrm{for} \: \Delta> \Delta_c, \: x>1, \\
g(x)&=&\frac{x^2}{\cosh^3 \left(\frac{1}{3} \cosh^{-1} x \right)} \left[1-\frac{x \tanh \left(\frac{1}{3} \cosh^{-1} x \right)}{3 \sqrt{x^2-1}}\right], \nonumber \\
& & \mathrm{for} \: \Delta< \Delta_c, \: x>1, \\
g(x)&=&\frac{x^2}{\cos^3 \left(\frac{1}{3} \cos^{-1} x \right)} \left[1-\frac{x \tan \left(\frac{1}{3} \cos^{-1} x \right)}{3 \sqrt{1-x^2}}\right], \nonumber \\
& & \mathrm{for} \: \Delta< \Delta_c, \: x<1.
\end{eqnarray} When $|E|>>L^{-1} \delta^{(z-1)\nu}$, these analytically obtained approximate crossover scaling functions have been compared with the numerically determined exact scaling function $\mathcal{R}$ in section~\ref{sec:numerics}. In this comparison, we have only adjusted an overall numerical prefactor for the analytical formula. A similar approximate expression for $\mathcal{C}$ is also found by substituting the explicit expressions of $\rho$ (in terms of $g(x)$) in Eq.~\ref{eqC}.

\begin{figure}[htbp]
 \begin{minipage}{\linewidth}
  \centering
  \includegraphics[width=0.7\linewidth,angle=-90]{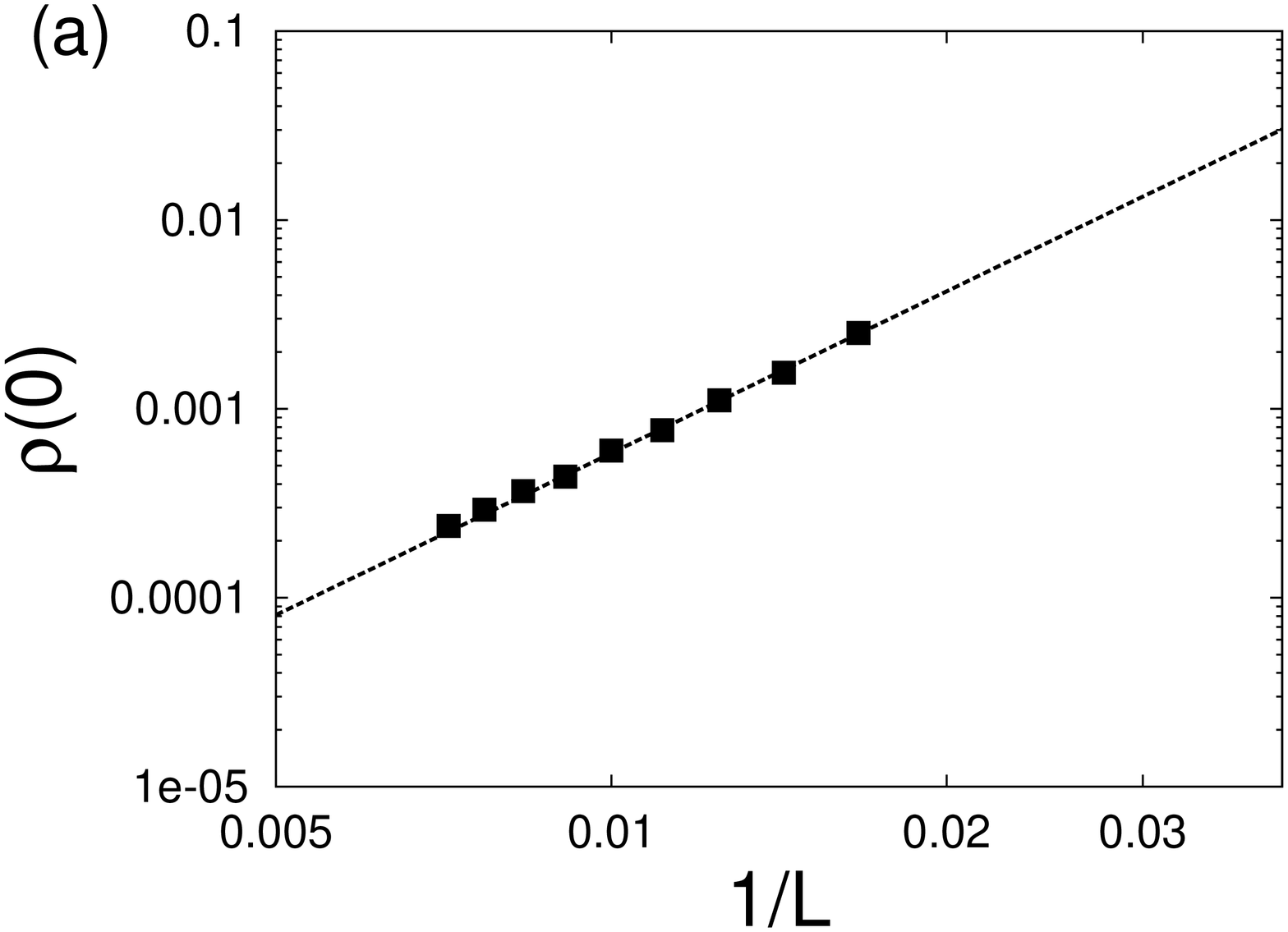}
\end{minipage}%
\newline
\begin{minipage}{\linewidth}
  \centering
  \includegraphics[width=0.7\linewidth,angle=-90]{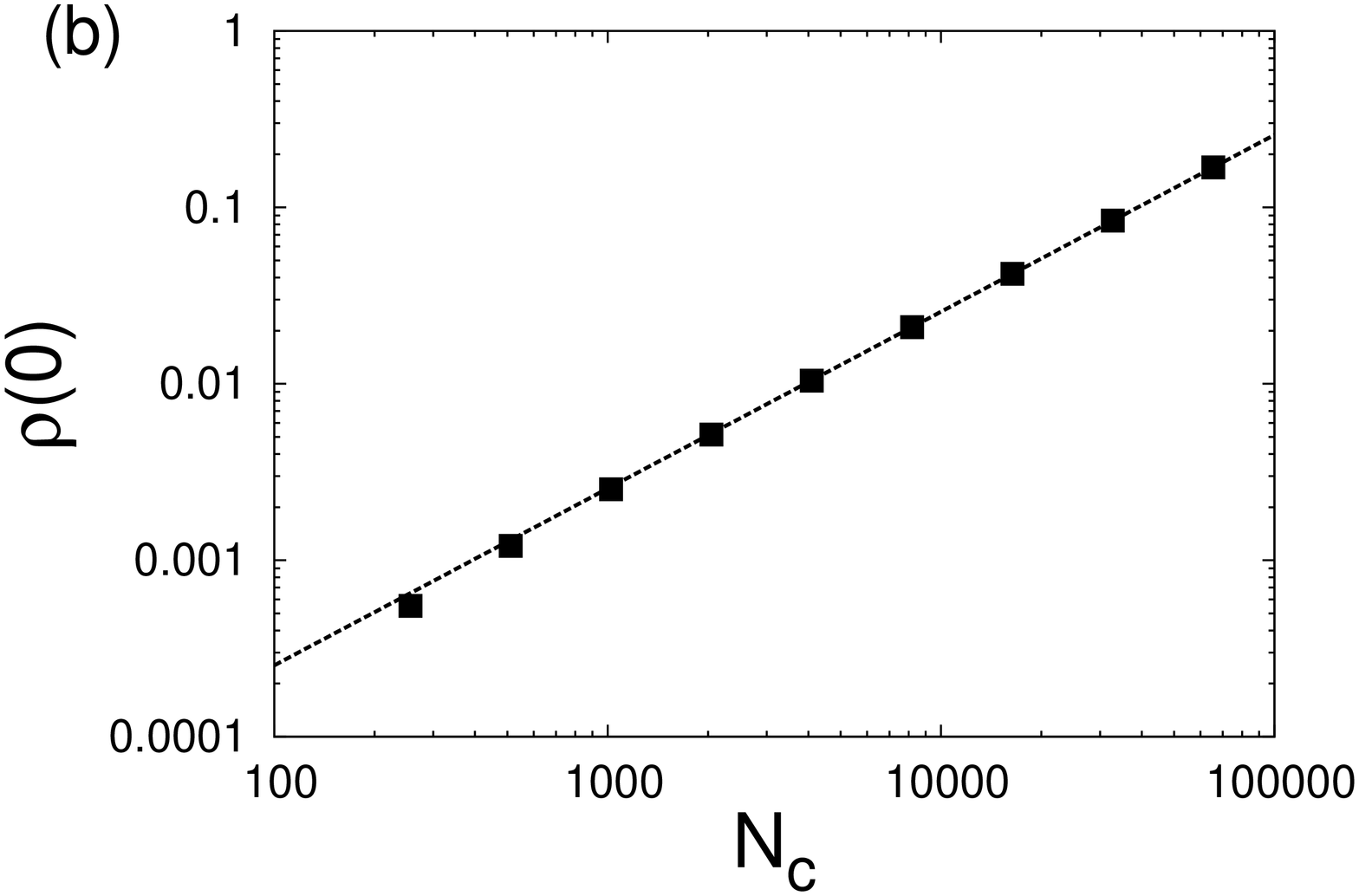}
\end{minipage}%
\caption{(a) (Color Online) Finite size scaling of $\rho(0,L,N_c)$ on a log-log scale, for $N_c =1028$ in the clean limit ($W=0$) yields a power law decay in system size $1/L^{2.85}$ consistent with $1/L^3$. (b) Scaling of  $\rho(0,L,N_c)$ on a log-log scale with $N_c$ for $L=60$ yielding a linear in $N_c$ scaling. Here $N_c$ is the number of terms kept in the Chebyshev expansion in the KPM numerics.}
 \label{fig:A1}
\end{figure}

\section{Numerical details}
\label{sec:appendix-C}
In this appendix we discuss the scaling and convergence of the zero energy DOS as a function of various KPM parameters. We begin with the clean limit $W=0$ and compute the zero energy density of states as a function of system size ranging from $L=60$ to $140$ in steps of $10$.
Generally in the thermodynamic limit we expect $\rho(0,L) \sim 1/L^2$.
\begin{figure}[htbp]
 \begin{minipage}{\linewidth}
  \centering
  \includegraphics[width=0.7\linewidth,angle=-90]{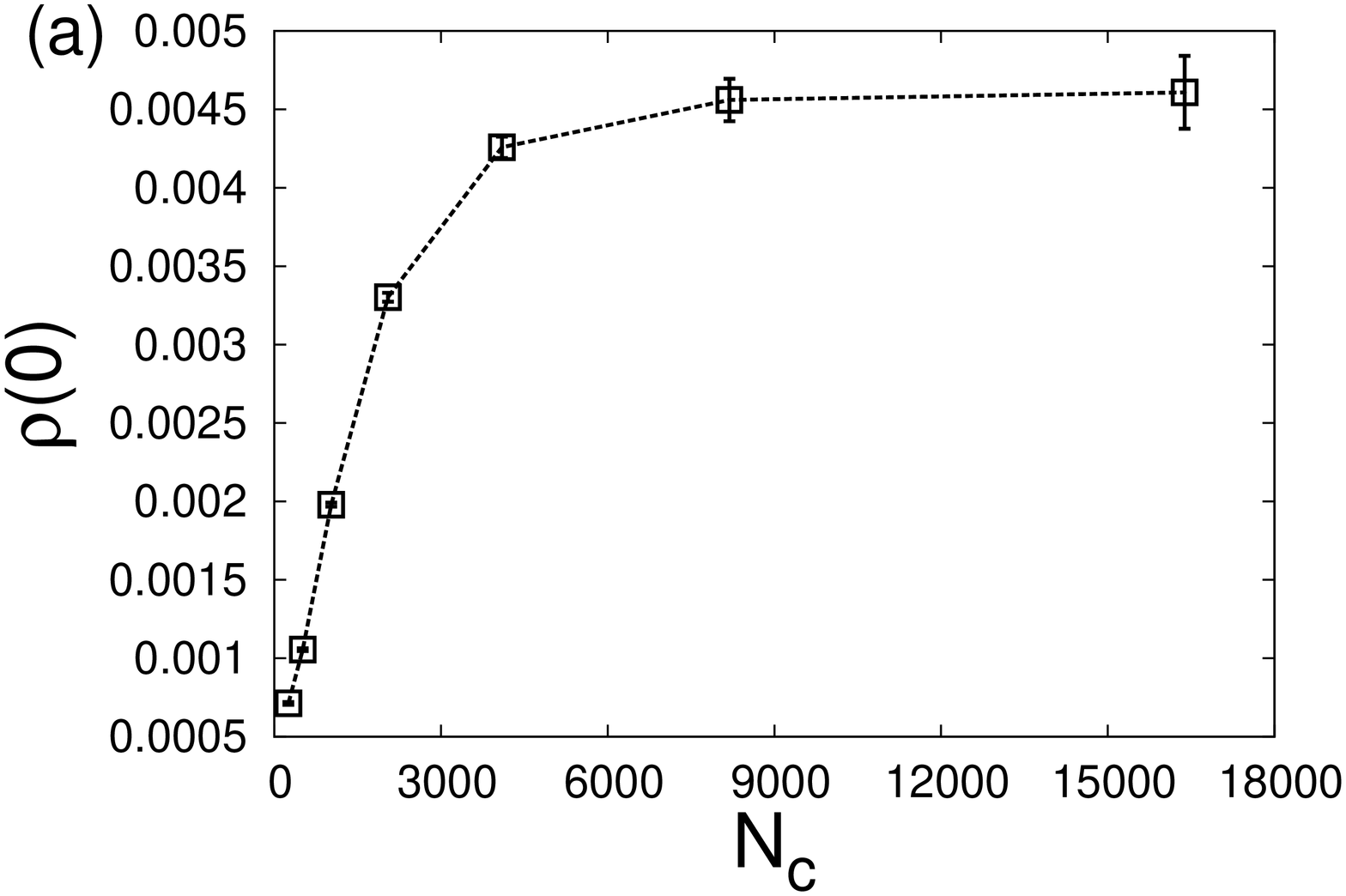}
\end{minipage}%
\newline
\begin{minipage}{\linewidth}
  \centering
  \includegraphics[width=0.7\linewidth,angle=-90]{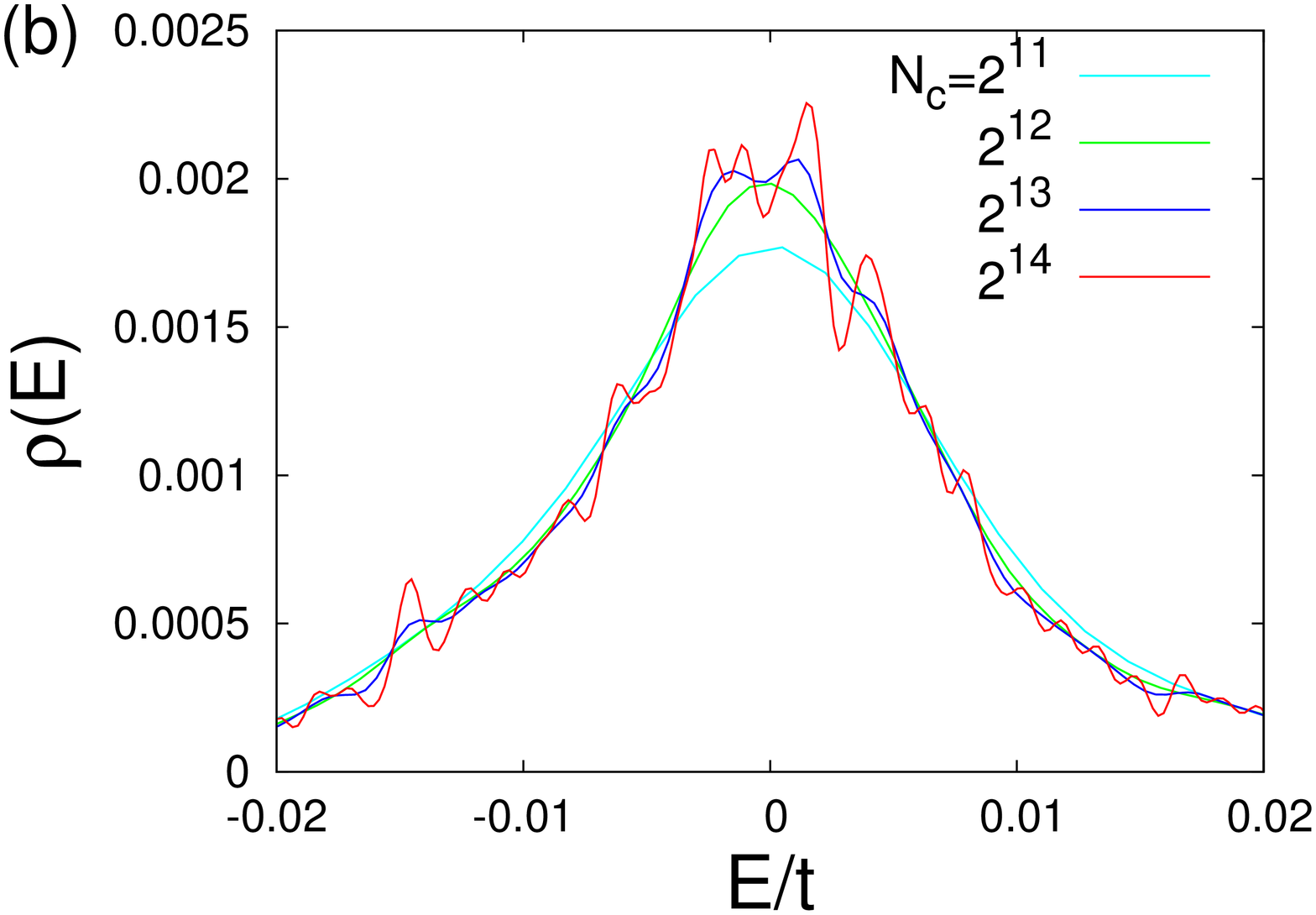}
\end{minipage}%
\caption{(Color Online) DOS for $L=60$ and $W=2.0 t < W_c$ (a) $N_c$ dependence of $\rho(0,L,N_c)$  converges in $N_c$ for $N_c \ge 4096$. (b) Low energy dependence of  $\rho(E,L,N_c)$ for various $N_c$ showing that the deviations in the DOS for larger $N_c$ in the SM phase is a result of the KPM beginning to resolve individual eigenstates. The low energy peak is a result of the 8 Dirac states that are there in the clean limit that have now been broadened by disorder.}
 \label{fig:A2}
\end{figure}
However, this behavior cannot be obtained within the KPM. Rather within the KPM we find $\rho(0,L) \sim 1/L^{2.85}$ consistent with a $1/L^3$ scaling [see Fig.~\ref{fig:A1}(a)].  This can be understood by considering the density of states with a Gaussian broadening factor $\sigma = \pi/N_c$, where $N_c$ is the number of terms kept in the Chebyshev expansion (as described in section~\ref{sec:numerics} A), as this is an excellent approximation to the effect of the Jackson kernel~\cite{Weisse-2006}. Since there is only an intensive number ($8$ corresponding to each cone) of states at zero energy the KPM yields
\begin{eqnarray}
\rho(0,L,N_c) &=& \sum_{n_x,n_y,n_z} \frac{\exp{\left[-\left(\frac{2\pi }{L\sigma}\right)^2(n_x^2+n_y^2+n_z^2)\right]}}{L^3\sqrt{2 \pi \sigma^2}}
\nonumber
\\
&\sim& \frac{1}{\sigma L^3}  = \frac{N_c}{\pi L^3}
\end{eqnarray}
where we have used $k_i = 2\pi n_i/L$.  We find $\rho(0,L,N_c)~\sim N_c^{1.02}$ reproducing the linear in $N_c$ scaling to a good degree of accuracy as shown in Fig.~\ref{fig:A1}(b). These results are obtained by considering the asymptotic behavior of the Jacobi theta function
\begin{equation}
\sum_{n=-\infty}^{\infty} e^{-x \pi n^2} = \theta_3(0, e^{-\pi x})=\vartheta(0, ix).
\end{equation}

\begin{figure}[htbp]
 \begin{minipage}{\linewidth}
  \centering
  \includegraphics[width=0.7\linewidth,angle=-90]{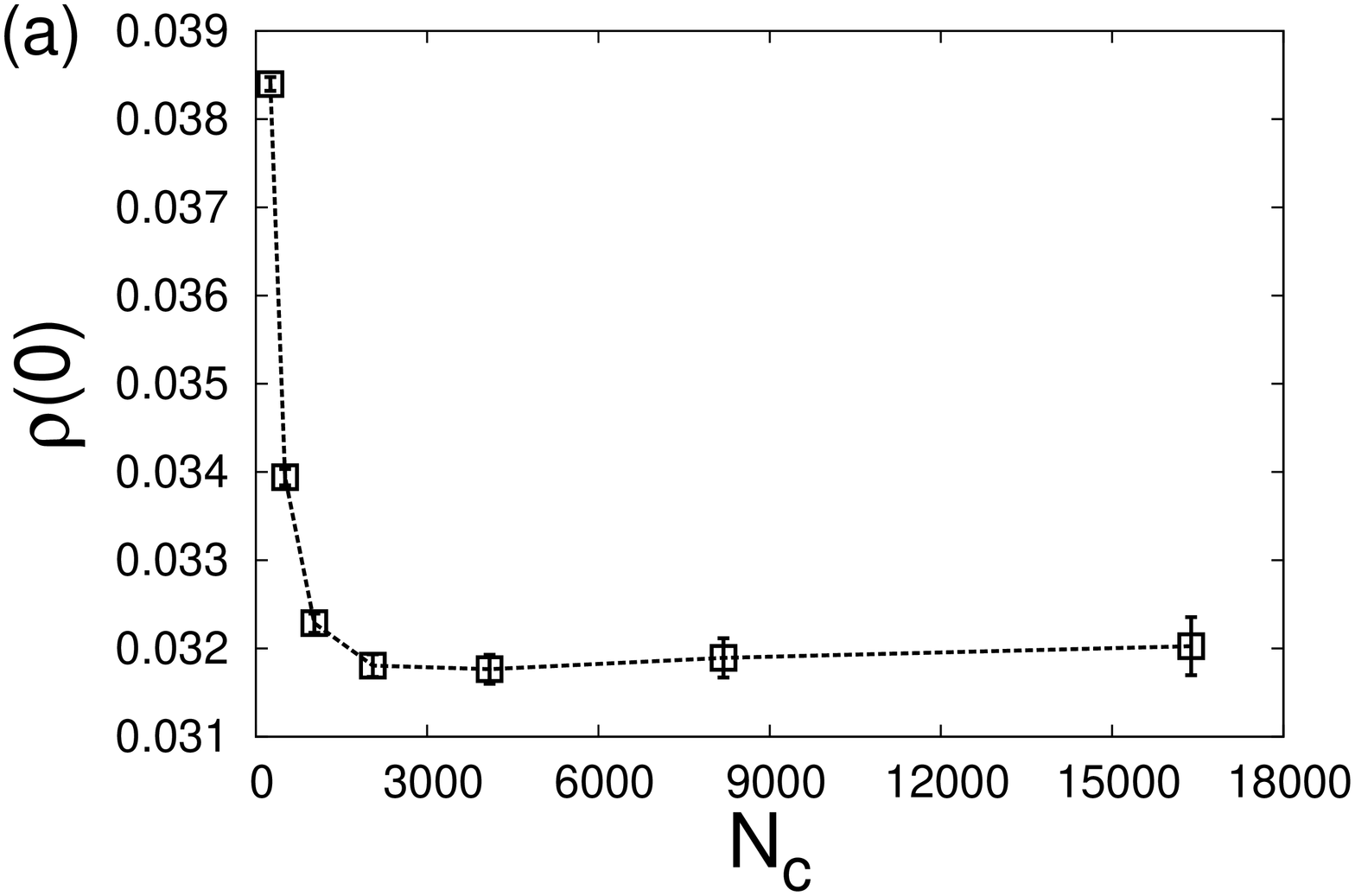}
\end{minipage}%
\newline
\begin{minipage}{\linewidth}
  \centering
  \includegraphics[width=0.7\linewidth,angle=-90]{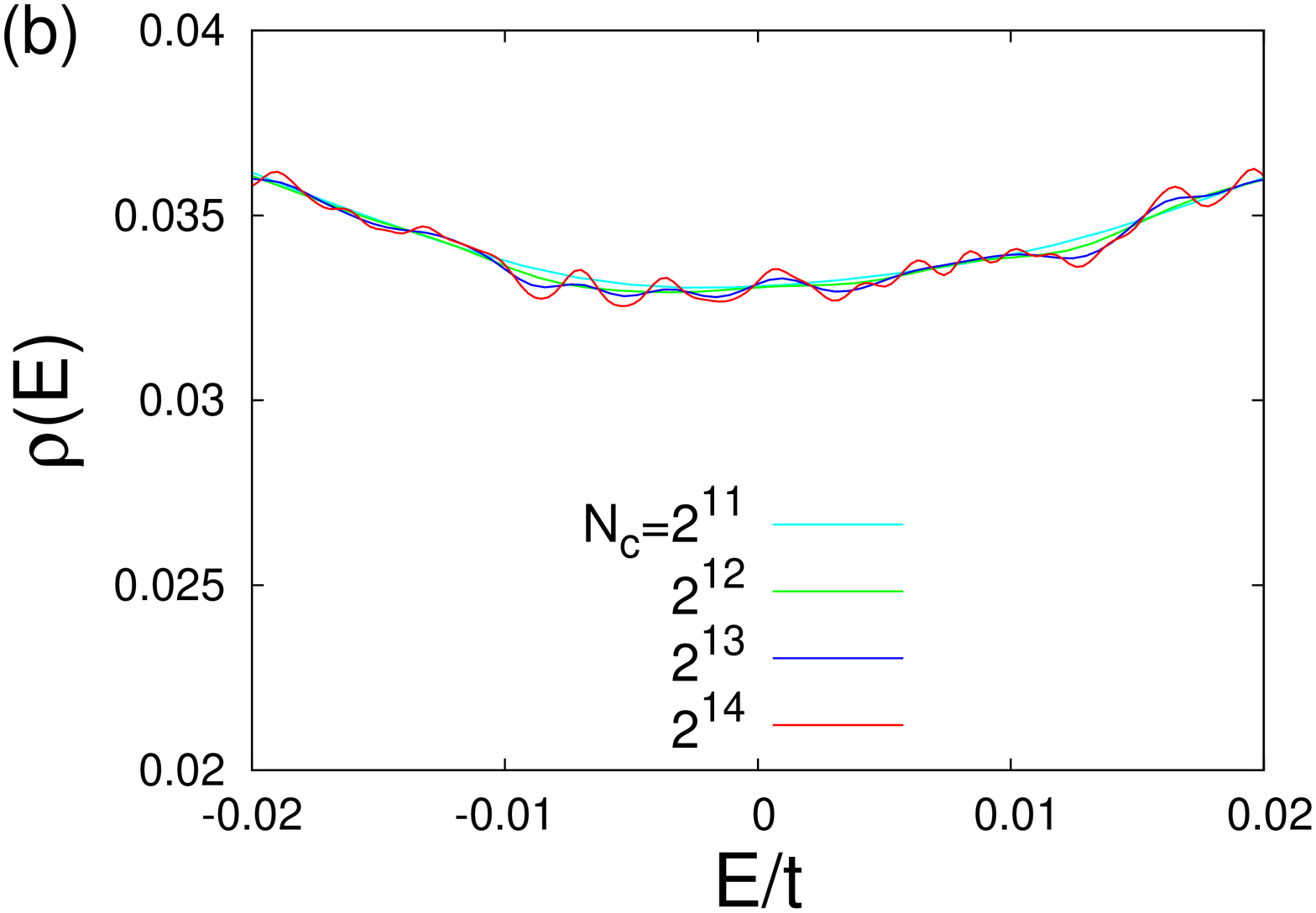}
\end{minipage}%
\caption{(Color Online) DOS for $L=60$ and $W=3.0 t> W_c$, (a) $N_c$ dependence of $\rho(0,L,N_c)$ converges in $N_c$ for $N_c \ge 4096$. (b) Low energy dependence of  $\rho(E,L,N_c)$ for various $N_c$ showing that there is a very weak dependence in the DM phase.}
 \label{fig:A3}
\end{figure}

Moving away from the clean limit by introducing disorder provides a natural broadening of the energy eigenvalues and thus changes the scaling with $N_c$. Focusing on the SM phase, we find that $\rho(0,L,N_c)\sim 1/L^3$ is essentially satisfied for weak disorder strengths. Even though we find a large deviation in the zero energy DOS as a function of $N_c$, it does converge for increasing $N_c$ as shown in Figs.~\ref{fig:A2} (a) and (b). The deviations in $\rho(0,L,N_c)$ are due to the KPM beginning to resolve the energy eigenstates.

\begin{figure}[htbp]
\centering
\includegraphics[width=0.7\linewidth,angle=-90]{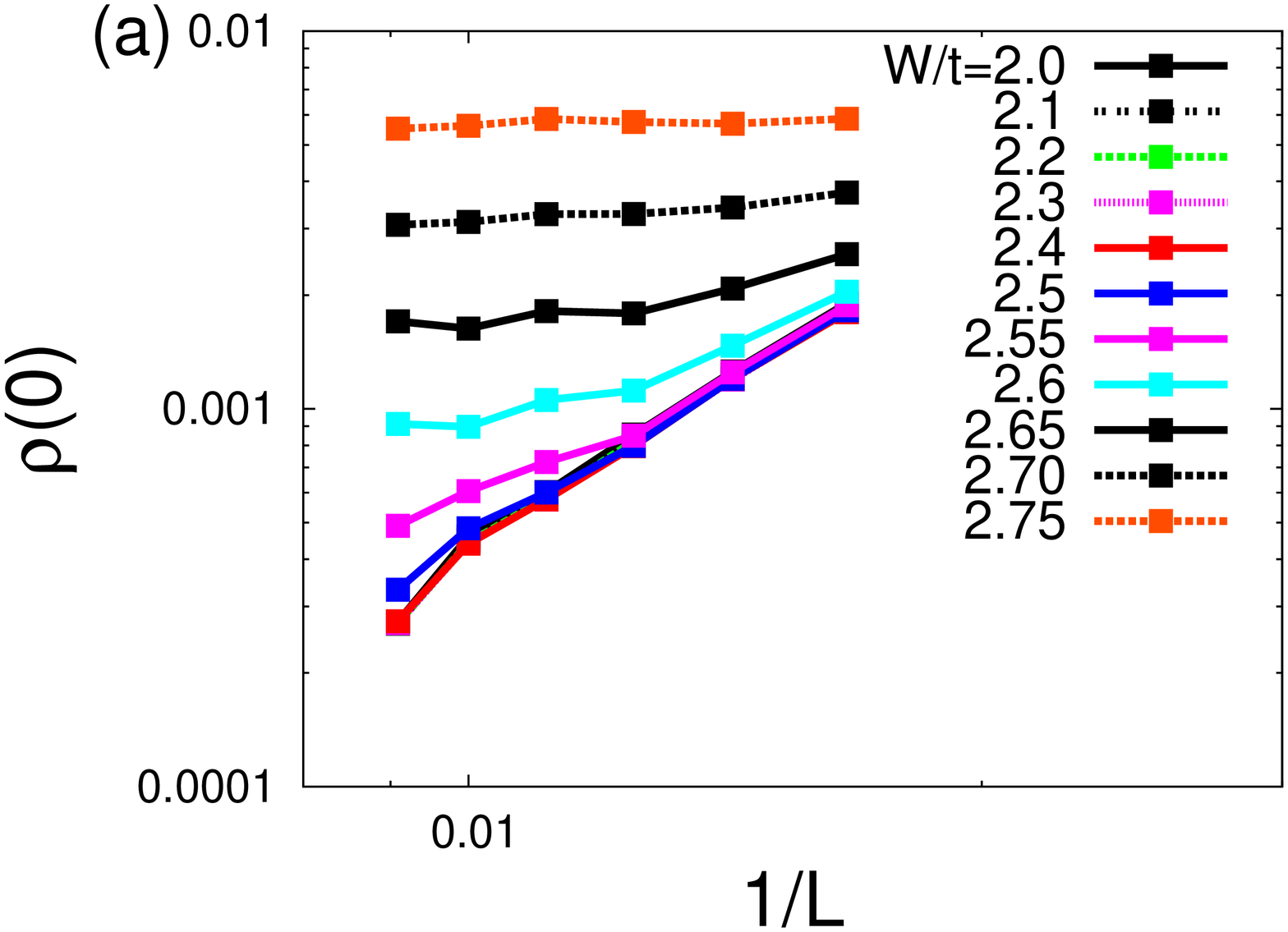}
\includegraphics[width=0.7\linewidth,angle=-90]{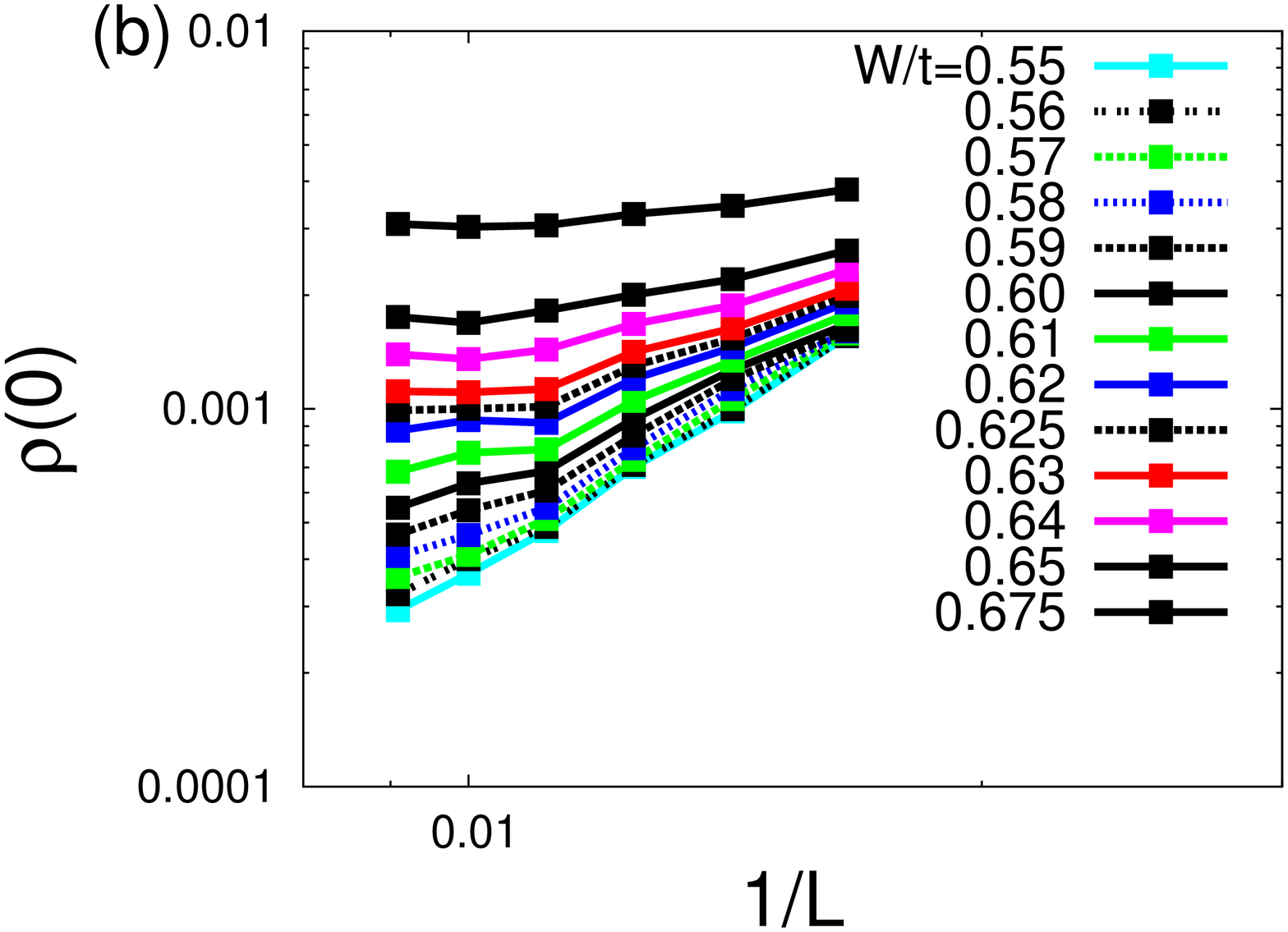}
\caption{(Color Online) DOS at zero energy $\rho(0)$ as a function of of the inverse system size $1/L$ on a log-log scale in both the SM and DM phases for potential disorder up to $L=110$. Results for the  box distribution (a) and the Gaussian distribution (b). In the SM phase  $\rho(0)$ goes to zero as the system size is increased, while it saturates in the DM phase, and at the QCP $\rho(0)$ is still decreasing for increasing $L$. In the DM phase the results are saturating in $1/L$.  }
\label{fig:A4}
\end{figure}

Inside the DM phase, $\rho(0,L,N_c)$ has a very weak dependence on $N_c$ [as shown in Figs.~\ref{fig:A3} (a) and (b)] and for sufficiently large system sizes,
converges in $L$ [see Figs.~\ref{fig:A4} (a) and (b)]. We conclude that in each phase choosing $N_c = 4096$ gives a converged estimate of $\rho(0)$, and therefore we use this value of $N_c$ to determine the critical exponents.

\newpage

\bibliography{DSM_QCP}

\end{document}